\documentclass[11pt]{article}

\usepackage[a4paper,
            left=2.2cm,
            right=2.2cm,
            top=3cm,
            bottom=3.5cm]{geometry}

\usepackage{graphicx}
\usepackage{subcaption}
\usepackage{placeins}
\usepackage{caption}
\usepackage{setspace}
\usepackage{titlesec}
\usepackage{enumitem}
\usepackage{abstract}

\titleformat{\section}
  {\large\bfseries}
  {\thesection}
  {1em}
  {}

\titleformat{\subsection}
  {\normalsize\bfseries}
  {\thesubsection}
  {1em}
  {}

\titleformat{\subsubsection}
  {\normalsize\itshape}
  {\thesubsubsection}
  {1em}
  {}

\title{\Large\bfseries
Materials Beyond Hamiltonian Limits\\[0.4em]
\large
Quantum Measurement as a Resource for Material Design\\[0.6em]
}

\author{Jochen Mannhart\\
Max Planck Institute for Solid State Research\\
70569 Stuttgart, Germany}

\date{February 24, 2026}

\begin{document}

\addtolength{\topskip}{-2cm}

\maketitle

\vspace{-1em} % space between author block and abstract

\begin{abstract}
Recent studies have identified materials and devices whose behavior lies beyond the scope of conventional electronic–structure theory. 
Such theories are formulated entirely in terms of Hamiltonian evolution and therefore describe only unitary dynamics and thus only a restricted class of quantum systems.

In contrast, electron systems that incorporate quantum measurement as an intrinsic dynamical element undergo Hamiltonian evolution interleaved with projection-induced state updates.
This unitary-projective dynamics breaks constraints imposed by purely unitary evolution and permits stochastic population transfer between symmetry-related transport channels, thereby enabling fundamentally new material functionalities.
This insight motivates the deliberate design of materials and devices that harness unitary–projective dynamics.

This article explores the foundations of unitary–projective electron dynamics and charts the resulting landscape of quantum materials and their functionalities. 
Model calculations demonstrate passive mesoscopic structures with intrinsic nonreciprocal single-electron transmission, materials exhibiting a novel category of magnetism, and possible platforms for energy harvesting and conversion with efficiencies that exceed the standard Carnot limit.

\vspace{1em}

\end{abstract}

\clearpage 
\section{Introduction}

Much of today’s thinking about quantum materials is based on the restrictive assumption that their properties are entirely determined by Hamiltonian–based physics. Within this established paradigm, electronic states evolve smoothly under unitary dynamics, the governing equations are linear, and, given the time-reversal symmetry of the microscopic processes, the resulting transport responses obey stringent reciprocity relations in the absence of explicit time-reversal-symmetry-breaking fields. These features naturally limit the range of responses and functions that quantum materials can exhibit.
Such constraints are evident in approaches such as the Kubo formalism \cite{Kubo1957, Kubo1966}, the Onsager relations \cite{Onsager1931a, Onsager1931b} and the Landauer–B\"uttiker formalism \cite{Landauer1957, Buttiker1986}.

However, strictly Hamiltonian evolution applies only to perfectly isolated quantum systems. Real quantum materials are open systems. They continuously exchange energy and information with unobserved degrees of freedom in the environment. Modeling such open systems (for an overview see, e.g., \cite{Breuer2002}) generally involves incorporating nonunitary elements, whether formulated as projection events, measurements, quantum collapses, or decoherence processes. These processes do not adhere to the assumptions of smooth, deterministic evolution that underlie standard Hamiltonian–based descriptions and induce discontinuous, stochastic state updates in the reduced electronic dynamics. As a result, the well-established constraints on material properties derived from pure Hamiltonian dynamics need not apply to electronic systems whose evolution includes projection events. 

Recent work \cite{Mannhart2018, Bredol2021b} has highlighted the role of projection processes in lifting Hamiltonian constraints for material design. This work has also explored how to engineer the combined action of Hamiltonian dynamics and projective state evolution to realize desired material functionalities. The considered projection processes generate state updates that violate detailed balance; thus, such systems operate in non-equilibrium steady states rather than relaxing to thermodynamic equilibrium.

Throughout this work, the term \textit{thermodynamic equilibrium} is used in the standard statistical-mechanical sense of Tolman and of Landau and Lifshitz \cite{Tolman1938, Landau1980}: a state with time-independent macroscopic observables and a stationary probability distribution that obeys detailed balance and is associated with maximum entropy under the conserved quantities.

Importantly, while this definition specifies the properties a state must possess to qualify as thermodynamic equilibrium, it does not guarantee that such a state is dynamically reachable or stable for a given system. In particular, if the reduced dynamics violates detailed balance, as in the systems considered here, thermodynamic equilibrium is dynamically inaccessible.
Instead, such systems may stabilize in dynamically selected non-equilibrium steady states. Because these states maximize entropy only within the subset of configurations dynamically accessible through the reduced evolution, the concepts, rules, and constraints of equilibrium thermodynamics are not generally applicable to them. As we show below, the systems examined in this study fall precisely into this category.

Consequently, these materials and devices lie outside the domains of both Hamiltonian transport theory and the framework of equilibrium thermodynamics. This opens a large design space for novel quantum materials and functionalities, which we explore in this article.

We will explain the scientific foundations of materials and devices governed by tailored combinations of Hamiltonian dynamics and projective state evolution, and outline the corresponding design principles. 
Such systems can be realized in any spatial dimension and across a wide range of material classes, including complex oxides, organic compounds, and metamaterial architectures. 
For illustrative purposes, we will discuss materials that exhibit nonreciprocal transmission probabilities for single-electron transport, effectively functioning as intrinsic, passive single-electron valves, as well as materials that display a novel form of magnetism generated by the combined action of thermal fluctuations and projection processes. Finally, we will present proposals for materials that enable thermoelectric energy conversion with super-Carnot efficiencies.

It is important to emphasize the scope and limitations of the present work. While the phenomenological and model-based approaches employed here reveal internally consistent behavior and robust physical mechanisms, a fully microscopic, first-principles theory of electron dynamics in open quantum systems that treats projection processes on equal footing with Hamiltonian evolution remains lacking. Thus, the results presented here are not intended to be quantitatively predictive or material-specific, but rather, they are preliminary, mechanism-based analyses that identify the conditions under which combined unitary–projective dynamics can relax conventional constraints and enable qualitatively new functionalities.

Throughout this work, the term \textit{constraints} refers to restrictions on admissible transport and response properties, such as reciprocity relations and linear-response bounds, that follow from the closed, unitary Hamiltonian dynamics of the reduced system. This definition does not apply to constraints in the sense of Dirac’s theory of constrained Hamiltonian systems or to kinematic reductions of the Hilbert space.

The paper begins by recapitulating the two fundamental ways in which quantum states may evolve under the basic axioms of quantum mechanics. Then, it shows how combining them enables quantum materials with new functional properties. It then presents concrete examples of how such materials can be designed and the properties they can exhibit. Finally, it concludes with a general discussion and outlook.

\section{Evolution of Quantum States}
\label{sec:Evolution_of_States}

Following an early formulation by Paul\,A.\,M. Dirac \cite{Dirac1930}, John von\,Neumann explicitly identified two distinct forms of quantum-state evolution: unitary evolution generated by Hamiltonian dynamics and state reduction through projection processes \cite{vonNeumann1932}. This distinction has remained central to the standard formulation of quantum mechanics. Indeed, these two forms of evolution are codified as axioms or postulates in the conventional mathematical formulation of quantum mechanics (see, e.g., \cite{Cohen1977, Shankar1994}). A brief comparison of the two state evolution paradigms is provided in Tab.~\ref{tab:Tab_A}.

\FloatBarrier

\begin{table}[t]
\centering
\renewcommand{\arraystretch}{1.2}

\setlength{\fboxsep}{7mm}
\setlength{\fboxrule}{0.6pt}

\newlength{\tabw}
\setlength{\tabw}{\dimexpr 0.46\textwidth + 0.1 cm + 0.46\textwidth\relax}

% Make caption width match the table width (optional but nice)
\captionsetup{margin={0 mm,-3.5 mm}}

\fbox{%
\begin{minipage}{\tabw}
\vspace{-3mm}
\raggedright
{\Large\bfseries Evolutions of Quantum States: Characteristic Features\par}\vspace{10pt}
{\footnotesize\raggedright\sloppy

\begin{tabular}{p{0.46\textwidth} @{\hspace{0.6 cm}} p{0.495\textwidth}}

\large\textbf{Unitary Evolution} & \large\textbf{Projective Evolution}\\[-2pt]
\textit{Coherent dynamics} & \textit{Measurement-driven (e.g., decoherence, collapse)}\\[-2pt]
\textit{Isolated quantum system} & \textit{Open quantum system}\\[-2pt]

\begin{itemize}[leftmargin=*, itemsep=2pt, topsep=0pt]
  \item Hamiltonian equations, e.g., Schr\"odinger, Dirac.
  \item Linear, deterministic.
  \item Information preserved by unitary evolution.
  \item Evolution is continuous as a function of time.
  \item Reversible microscopic dynamics,\newline
  time-reversal symmetry for time-reversal-invariant Hamiltonians.
  \item System properties constrained by Onsa\-ger/Kubo/Landauer–Büttiker relations under equilibrium and linear-response conditions.
  \item Thermodynamic equilibrium is accessible and stable, so equilibrium constraints apply \newline 
  (standard textbook behavior).
  
\end{itemize}
&
\begin{itemize}[leftmargin=*, itemsep=2pt, topsep=0pt]
  \item Born rule, von\,Neumann–L\"uders state-update postulate.
  \item Nonlinear and stochastic conditioned state updates.
  \item Information exchanged with the environment (decoherence models) or stochastically reset by collapse dynamics (collapse models).
  \item Evolution can be discontinuous as a function of time.
  \item Reduced dynamics generally irreversible,\newline 
        detailed balance need not hold in the reduced dynamics.
  \item Onsager, Kubo, Landauer--B\"uttiker constraints are not always applicable, as their validity depends on equilibrium, detailed balance, and linear response.
  \item Thermodynamic equilibrium may be dynamically inaccessible, so equilibrium constraints need not apply.

\end{itemize}
\end{tabular}
}
\vspace{-3mm}
\end{minipage}
}

\caption{\textbf{Overview of quantum-state evolution.}
This figure illustrates the characteristic features of the two fundamental forms of quantum-state evolution, unitary (Hamiltonian) evolution and projective (measurement-induced) evolution, as defined by the basic axioms of quantum mechanics. The measurement process underlying projective evolution arises intrinsically from the system’s dynamics, either through coupling to environmental degrees of freedom (as in decoherence-based descriptions) or through intrinsic collapse mechanisms. Therefore, projective evolution does not require an external measurement apparatus or an observer. Listed are the features central to the electron dynamics addressed in this manuscript.}
\label{tab:Tab_A}
\end{table}

\subsection{Evolution by Hamiltonian Dynamics and Its Limits}
\label{subsec:Hamiltonian_dynamics}

We begin by reviewing the standard approach to modeling quantum-state evolution, namely modeling unitary evolution generated by Hamiltonian dynamics for isolated systems, as commonly implemented within the highly successful framework of electronic–structure theory. In this paradigm, the properties of an electronic system are determined by specifying an appropriate Hamiltonian and solving the corresponding Schr\"odinger or Dirac equation. The unitary time evolution of the electronic states follows directly from the Hamiltonian, and solving it yields the ground state as well as the spectrum of excited states.

While this procedure is straightforward in principle, it becomes computationally demanding in practice because the numerical effort grows rapidly with the number of electrons. This is due to the fact that the dimension of the underlying Hilbert space grows exponentially with system size. As early as 1929, P.\,A.\,M.\;Dirac emphasized this point, noting that \textit{“the underlying physical laws necessary for the mathematical theory of a large part of physics and the whole of chemistry are thus completely known, and the difficulty is only that the exact application of these laws leads to equations much too complicated to be soluble”} \cite{Dirac1929}.

Nevertheless, by incorporating the ionic lattice potential, electron–electron interactions, and other relevant contributions into the Hamiltonian, a wide range of electronic phenomena central to modern quantum materials can already be described within Hamiltonian-based, unitary quantum mechanics. Examples include spin–orbit effects, topological phases, strong electronic correlations, and interface phenomena such as two-dimensional electron systems.

To address the numerical challenges posed by many-electron Hamiltonians, a broad variety of theoretical and computational methods have been developed and continue to be refined.
These include density functional theory \cite{Hohenberg1964, Kohn1965},
dynamical mean-field theory \cite{Georges1992},
density-matrix renormalization group and related renormalization-group approaches \cite{White1992, Schollwock2011, Metzner2012},
quantum Monte Carlo techniques \cite{Ceperley1980, Foulkes2001},
Green’s-function methods \cite{Hedin1965, Fetter1971},
and advanced quantum-chemistry approaches \cite{Cizek1966, Bartlett2007, Booth2009}.
These methods differ substantially in their formulation: some solve effective eigenvalue problems, while others use propagators or stochastic sampling. All of these methods are powerful tools for describing the unitary dynamics of a Hamiltonian-based physical system.

This work does not suggest improvements to these methods, which are highly successful and well-established in their respective domains. 
Rather, we emphasize a complementary perspective. By augmenting exclusively Hamiltonian-based approaches with projection-driven dynamics, the applicability of electronic-structure descriptions can be extended to regimes inaccessible under purely unitary evolution. 
Because Hamiltonian-based approaches are linear with respect to the quantum state (the Schr\"odinger equation is linear in $\psi$) and describe unitary dynamics, even though they may give rise to nonlinear equations for observables or effective fields, the class of material behaviors and functionalities they can capture is fundamentally constrained.

In standard electronic–structure theory, the quantum state dynamics is necessarily treated as smooth and deterministic. Probabilistic measurement outcomes only arise when projective state updates are included. The Hamiltonian is Hermitian and often time-reversal symmetric unless time-reversal symmetry is explicitly or spontaneously broken, for example, by magnetic fields or the choice of a symmetry-broken ground state. This ensures real eigenvalues and imposes additional constraints on the allowed electronic responses. While real spectra may also arise in certain non-Hermitian settings \cite{Bender1998}, the governing equations remain linear. Consequently, any physical state can be expressed as a superposition of eigenstates, and observables are determined by the amplitudes and relative phases of these components.

The structural properties of Hamiltonian-based quantum mechanics—linearity, Hermiticity, unitary time evolution, and, in many cases, time-reversal symmetry—underlie the central constraints on material properties encoded in the Onsager relations \cite{Onsager1931a, Onsager1931b}, the Kubo linear-response theory \cite{Kubo1957, Kubo1966}, the fluctuation–dissipation theorem \cite{Callen1951}, the Landauer–B\"uttiker approach to electronic transport \cite{Landauer1957, Buttiker1986}, and the many models built upon them.

In the Onsager framework, microreversibility and the linearity of the underlying transport equations ensure that transport coefficients are symmetric under the exchange of driving forces and responses, provided that the Hamiltonian is time-reversal symmetric. Kubo’s linear response theory is formulated under the assumption that small perturbations generate proportional responses governed by unitary evolution under a Hermitian Hamiltonian. The resulting response functions obey strict reciprocity relations when time-reversal symmetry is preserved. 

Likewise, the Landauer–B\"uttiker model \cite{Landauer1957, Buttiker1986} relies on coherent, elastic, and unitary scattering processes. Because the underlying scattering matrix is unitary and typically constrained by time-reversal symmetry, the permitted transmission probabilities obey reciprocity conditions that constrain the possible current–voltage characteristics of mesoscopic conductors. 

Inelastic or dephasing effects are often incorporated into this model through B\"uttiker probes \cite{Buttiker1986}, fictitious terminals that mimic phase breaking by coupling the conductor to an electron reservoir. In this construction, the enlarged system (the conductor plus the probe leads) is still described by a globally unitary scattering matrix, and reciprocity remains intact as long as no time-reversal-symmetry-breaking bias or field is introduced. 
Imposing a zero-net-current condition at the probe ensures that electrons entering the probe are replaced by reservoir electrons. Because these have random phases, phase coherence is erased in the reduced description. However, B\"uttiker probes implement dephasing without invoking projective (von\,Neumann-L\"uders type) state updates. By construction, the Landauer–Büttiker formalism is therefore restricted to stationary states, precluding the dynamical effects of interest in later sections of this paper.

These formalisms are widely regarded as defining the fundamental boundaries within which quantum materials and electronic devices must operate. The Onsager relations, the Kubo formalism, and the Landauer–B\"uttiker approach not only provide powerful tools for calculating transport and response properties, but also serve as guiding principles that delineate permissible behavior within quantum transport. From this perspective, the derived reciprocity relations, linear-response assumptions, and unitary scattering constraints are regarded as fundamental properties of electronic materials in general, imposing stringent limits on phenomena such as directional conductance, nonreciprocal transport, and response functions.

Real materials, of course, are coupled to their environments and subject to thermal fluctuations. Standard electronic-structure theory typically incorporates finite-temperature effects by thermally populating Hamiltonian eigenstates or quasiparticle excitations. Alternatively, temperature-dependent scattering rates and lifetime broadening are introduced through self-energies (e.g., from electron-phonon coupling), followed by thermal averaging of observables. The resulting temperature-dependent occupation factors smear spectral features and modify the computed transport responses.

\begin{figure}[!p]
% \vspace*{-1.8cm}
\centering
\captionsetup{width=0.9\textwidth}
\includegraphics[width=0.9\textwidth]{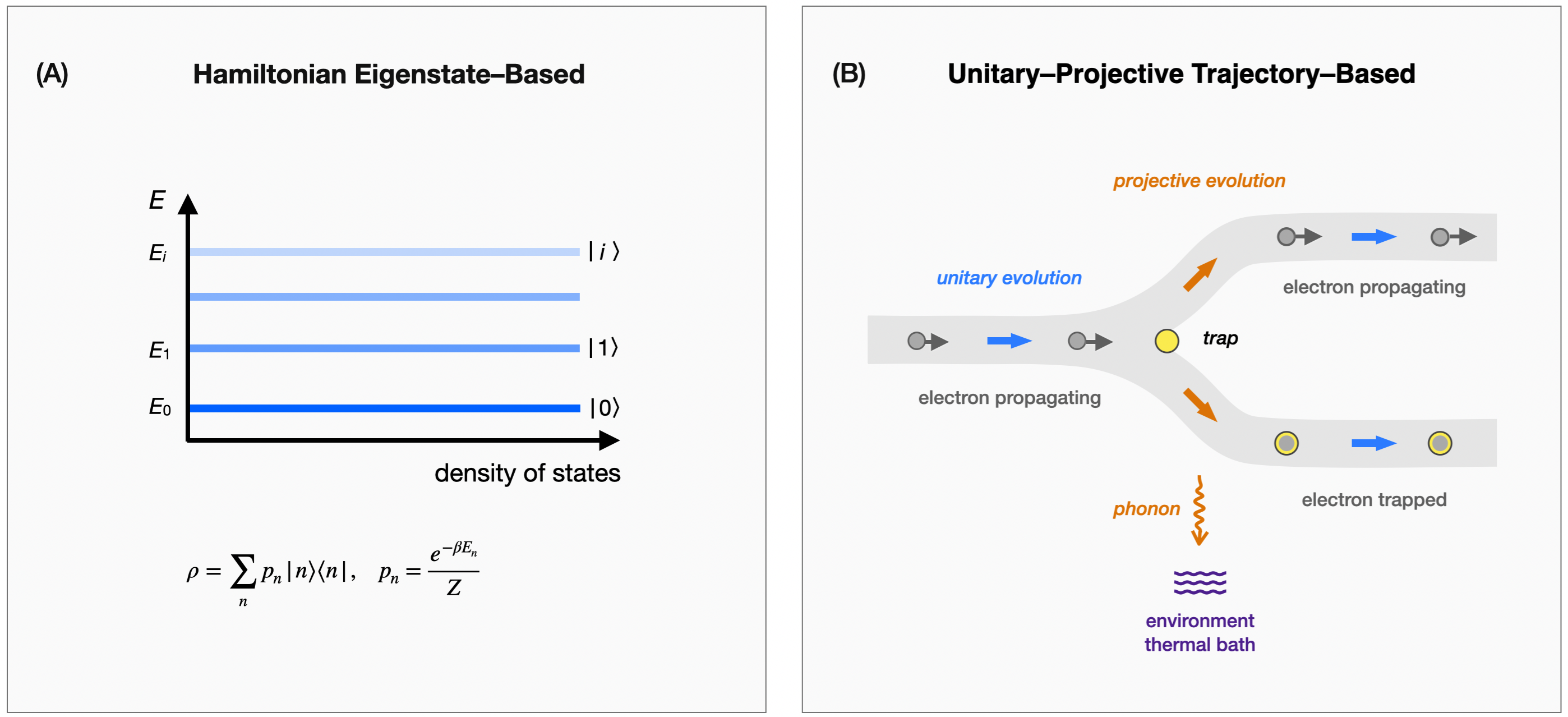}
\vspace{5 mm}
\caption{\textbf{Descriptions of quantum states at finite temperatures.} \\
\textbf{(A)}  
Schematic illustration of a conventional Hamiltonian-based description of electronic systems at a finite temperature. This framework applies in regimes where electronic coherence lengths and mean free paths are large compared to microscopic length scales, such that electronic states are well described by stationary energy eigenstates of an underlying Hamiltonian. Solving the Schr\"odinger or Dirac equation yields energy eigenstates $|n\rangle$ and eigenvalues $E_n$ of an underlying Hamiltonian. Finite-temperature effects are  incorporated by assigning thermal occupation probabilities (indicated by the saturation of the blue color representing the eigenstates in the figure) to these eigenstates or to corresponding quasiparticle states, resulting in a thermal ensemble described by a density matrix $\rho$.
In more elaborate treatments, temperature is incorporated through thermally induced scattering and lifetime broadening encoded in self-energies, or through thermal sampling of ionic configurations. In all cases, the description remains grounded in an underlying Hamiltonian eigenstate or quasiparticle structure, which provides the basis states populated by the electrons.\\
\textbf{(B)} Illustration of unitary–projective evolution of an electron trajectory encountering a trap site (an inelastic scattering center), sketched in a quantum-trajectory setting. This description is appropriate in the  regime where inelastic scattering events occur on timescales comparable to the electron’s traversal time through the system. An electron incident from the left propagates ballistically under unitary Hamiltonian dynamics until it encounters a trap. There, coupling to a phonon bath enables a projective event: the electron is either captured by the trap (downward branch) or it continues propagating in a null-measurement branch (upward arrow). In both cases, the subsequent evolution is again unitary until the electron encounters another trap or is released from the trap by a thermal fluctuation, allowing the process to repeat. Here, temperature governs the trap occupancy, the detrapping rate, and the energy distribution of the states occupied after release. The resulting trajectories are not energy eigenstates, and the associated thermal ensemble differs qualitatively from the thermal ensemble of Hamiltonian eigenstates as illustrated in (A).}

\label{fig:Fig_B}
\end{figure}

Importantly, as Fig.\,\ref{fig:Fig_B}(A) schematically summarizes, this procedure leaves the core assumption of Hamiltonian-based theory intact: the electronic system is meaningfully described in terms of coherent evolution within the eigenstate or quasiparticle structure of a Hamiltonian. This structure provides the basis states populated by electrons. 

However, if projection events hinder the electronic system from evolving coherently, this assumption breaks down in a fundamental way. Mathematically, the eigenstates still exist, of course, but environmental monitoring at a finite temperature prevents electrons from maintaining the coherent superpositions required to follow those eigenstates beyond the decoherence timescale. This situation differs qualitatively from ordinary quasiparticle decay. In that case, the eigenstate description remains operational, and coherence only gradually diminishes. 

In contrast, projection-like dynamics drives the system towards statistical mixtures that cannot be meaningfully interpreted within a Hamiltonian eigenbasis (see Fig.\,\ref{fig:Fig_B}(B)). A description based on Hamiltonian eigenstates then becomes invalid. Correspondingly, band structures and band diagrams cease to provide an adequate framework for describing the system’s dynamics and steady states.
Further, phenomena that rely on coherent band evolution and adiabatic transport, such as certain topological protection mechanisms, may no longer apply.

This exposes a critical issue: the conventional approaches, whether implemented through eigenstate\allowbreak-based methods or through Hamiltonian-derived formalisms, presuppose conditions that are not valid under environmentally induced projection. Consequently, these approaches exclude \textit{by assumption} entire classes of possible electronic behavior that lie beyond the reciprocity, linear-response, and unitarity constraints imposed by the Onsager–Kubo–Landauer–B\"uttiker-type theories. 
These standard frameworks assume Hamiltonian or weakly dissipative reduced dynamics that admits a linear-response description around a stationary reference state. While they can accommodate broken time-reversal symmetry and coupling to reservoirs, they do not incorporate stochastic projection events that violate detailed balance in the reduced dynamics. 
When a subsystem undergoes stochastic state updates induced by projection, driven by the environment, for example, those assumptions no longer hold, and the system’s properties may fall outside the limits of Hamiltonian electronic-structure theory.

By \textit{Hamiltonian limits} or \textit{Hamiltonian constraints} we refer to restrictions that apply to electronic systems governed by closed, unitary, and microreversible dynamics, such as two-terminal linear-response transport described by unitary scattering. This terminology does not question the existence of an underlying Hamiltonian for the combined system and its environment as used in standard decoherence theory. Rather, it addresses the breakdown of a purely Hamiltonian description of the reduced electronic dynamics, where present.

\FloatBarrier

%\clearpage

\subsection{Evolution by Projection Processes}
\label{subsec:Projection_dynamics}

Within the mathematical framework of quantum mechanics, the evolution of quantum systems coupled to an environment or to a continuum of states may involve projection processes. In the widely used open-system descriptions, the leakage of information into environmental degrees of freedom is generally sufficient to induce decoherence. Upon tracing out the environment, the resulting reduced dynamics effectively becomes projection-like, even when the system is in thermal equilibrium with its environment. Mathematically, projection processes constitute a mode of quantum-state evolution that is fundamentally distinct from Hamiltonian dynamics. 

The physical interpretation of projection or quantum measurement processes has long been debated with respect to the foundations of quantum mechanics, and has given rise to a wide range of conceptual viewpoints (see, e.g., \cite{Wheeler1983, Greenberger2009}). The spectrum of interpretations includes approaches in which the combined system-environment evolves unitarily under Hamiltonian dynamics and models invoking genuine physical collapse processes, among others.
Despite these interpretational differences, the mathematical description of projective processes within the standard formalism is precise, rigorous, and remarkably successful. Projective measurements are represented by well-defined projection operators acting on the system’s state. Outcome probabilities are given by the Born rule \cite{Born1926}, and probabilistic post-measurement state updates are prescribed by the von\,Neumann--L\"uders measurement postulate \cite{vonNeumann1932, Lueders1951}. 

The Born rule assigns probabilities to measurement outcomes. For a system in state $|\psi\rangle$, the probability of obtaining outcome $i$, associated with a projection operator $P_i$, is given by $p_i = \langle \psi | P_i | \psi \rangle$. Thus, the Born rule provides a mapping from quantum amplitudes to classical probabilities, but it does not, by itself, specify a state transformation or a dynamical update of the system.

The von\,Neumann--L\"uders measurement postulate, by contrast, specifies the conditional state update associated with a particular measurement outcome. Upon obtaining outcome $i$, the system state is updated according to
$|\psi\rangle \;\rightarrow\; {P_i|\psi\rangle}/{\sqrt{\langle \psi | P_i | \psi \rangle}} $. This is a nonlinear and stochastic transformation of the quantum state. 
Stochasticity in the reduced dynamics arises from the random selection of outcomes according to the Born probabilities, while nonlinearity and effective irreversibility arise from the conditioning and renormalization inherent in the post-measurement state update. The combination of probabilistic outcome selection and conditional state update renders projective processes stochastic, nonlinear, and effectively irreversible at the level of the reduced system dynamics.
While von\,Neumann introduced the general measurement framework \cite{vonNeumann1932}, L\"uders formulated the physically minimal projection rule for degenerate observables \cite{Lueders1951}. We therefore follow the usage of calling it "von\,Neumann–L\"uders projection postulate".

In this work, the term \textit{projection} refers precisely to such von\,Neumann--L\"uders-type conditional state updates acting on the reduced electronic subsystem, whether arising from explicit measurements, environment-induced decoherence, or effective collapse processes. The resulting dynamics therefore goes beyond purely unitary Hamiltonian evolution and plays a central role in enabling the unitary–projective functionalities discussed below.

In systems weakly coupled to continua of final states, the Born-rule transition probabilities are commonly computed by using Fermi's golden rule  \cite{Dirac1927, Fermi1931, Fermi1949}, which follows directly from the Born rule in the weak-coupling, continuum limit \cite{Cohen1992}. 
The Born rule has been applied in countless experimental settings and has never been found to fail. By providing a quantitative link between quantum amplitudes and classical probabilities, the Born rule is a cornerstone of modern quantum physics and quantum technologies.

Projection processes in the electronic (reduced) system break several structural features that constrain Hamiltonian-based dynamics: they generate effectively discontinuous state transitions, are intrinsically stochastic, induce irreversible information updates, and can break microreversibility in the reduced dynamics (see Fig.\,\ref{fig:Fig_B}(B) for an illustration). Moreover, the associated conditional state updates are nonlinear as a consequence of Born-rule probabilities combined with the required normalization of post-measurement states \cite{Breuer2002, Plenio1998}.
Consequently, the resulting dynamics falls squarely outside the regime in which reciprocity relations, linear-response behavior, and unitary scattering constraints, such as scattering-matrix unitarity arising from probability conservation, are derived and expected to hold.

With \textit{projection event} we refer to a localized, stochastic state update, arising either from coupling to environmental degrees of freedom or from intrinsic collapse dynamics, that irreversibly modifies the information content of the reduced system. As an illustrative example, we consider the capture of electrons at defects in crystalline lattices accompanied by phonon emission into a bath.
Although often regarded as mere sources of decoherence or measurement backaction, projective events can be deliberately engineered and harnessed as a non-exhaustible resource. When combined with Hamiltonian dynamics, projective events provide a powerful mechanism for crafting qualitatively new electronic dynamics.

\subsection{Evolution Under Unitary Hamiltonian Dynamics and Projection Processes}
In realistic settings, quantum states evolve under the combined action of coherent Hamiltonian dynamics and nonunitary, projection-like processes. As discussed in Sect.~\ref{subsec:Hamiltonian_dynamics}, information leakage to the environment erodes coherence; consequently, the standard eigenstate- or quasiparticle-based thermal-averaging approach therefore no longer provides an adequate description of the dynamics. 

This limitation signals the need for a framework that goes beyond Hamiltonian evolution. One such framework is the theory of open quantum systems \cite{Breuer2002}, in which the combined action of unitary evolution and decoherence can profoundly reshape electronic behavior. Alternatively, equivalent reduced dynamics can, for example, be described within objective collapse models. Because these descriptions are nominally equivalent for the reduced electronic dynamics, we adopt the open-quantum-system framework in the following as a representative formulation, for clarity of presentation.

Within this framework, a variety of theoretical approaches are used to describe the interplay between unitary evolution and projection processes, including quantum-trajectory methods, Lindblad or Redfield master equations, and non-equilibrium Green’s-function techniques (see, e.g., \cite{Breuer2002, Plenio1998, Carmichael1993}). These approaches have demonstrated that coupling to an environment can qualitatively modify transport and response properties.

One example is dephasing-assisted transport, where moderate decoherence restores conduction in systems that would otherwise be localized \cite{Haken1973, Rebentrost2009}. In nanoscale conductors, controlled dephasing can enhance current flow by mitigating interference-induced suppression of transport paths, an effect captured by B\"uttiker’s dephasing-probe model \cite{Buttiker1986}. 
Related mechanisms play a central role in specific material classes. In molecular and organic materials, phonon-assisted hopping driven by thermal lattice fluctuations enables efficient charge motion \cite{Baessler1993}. In biological light-harvesting complexes, environment-assisted exciton transport enhances energy-transfer efficiency under physiologically relevant noise conditions \cite{Engel2007, Plenio2008}. 
Dissipation can furthermore stabilize non-equilibrium quantum phases that have no analog in closed systems. Engineered loss and decoherence have been shown to generate topological steady states \cite{Bardyn2013} and to support driven--dissipative condensates, such as exciton--polariton fluids in semiconductor microcavities \cite{Kasprzak2006}. 
A prominent example of the deliberate combination of unitary evolution with controlled projection processes to enhance functionality is found in quantum error correction. In  error correction protocols such as the Shor code \cite{Shor1995}, logical qubits are stabilized by repeatedly measuring error syndromes, with the measurement outcomes used to identify and suppress errors.
Together with the preceding examples, this illustrates that, when combined with projection processes, coherent Hamiltonian dynamics can give rise to functionalities that are fundamentally inaccessible under purely unitary evolution.

\begin{figure}[!p]
\vspace*{-1.8cm}
\centering
\captionsetup{width=0.9\textwidth}
\includegraphics[width=0.9\textwidth]{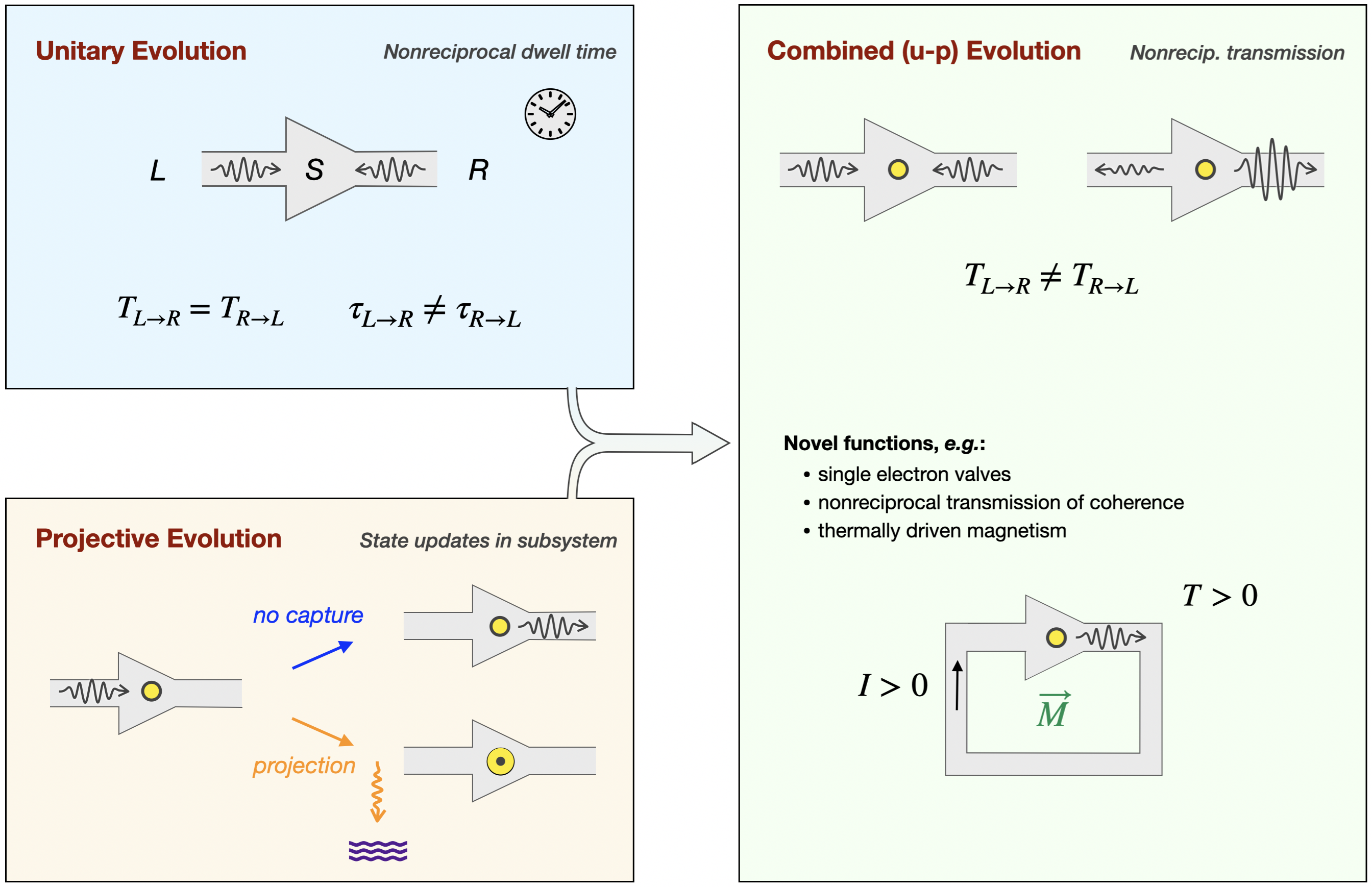}
\vspace{5 mm}
\caption{\textbf{Design principle of unitary–projective devices with emergent non-equilibrium functionality.}
The central idea is to combine coherent (Hamiltonian) but symmetry-broken propagation with localized projective state updates. This converts internal time asymmetry into directional transport and non-equilibrium order.\\
\textit{Unitary Hamiltonian evolution (top left):} Although coherent unitary dynamics in a spatially asymmetric structure can produce direction-dependent dwell times $\tau_{L \to R}$ and $\tau_{R \to L}$, the absence of projections and explicit time-reversal-symmetry breaking ensures that the scattering matrix $S$ remains unitary. Consequently, the forward and backward transmission probabilities $T_{L \to R}$ and $T_{R \to L}$ are identical, in accordance with microreversibility and Onsager reciprocity.\\ 
\textit{Projective, stochastic state updates (bottom left):} Projection-like processes export phase information from the device (subsystem) to the environment, giving rise to nonunitary reduced dynamics and, in non-equilibrium settings, the breakdown of detailed balance. Such behavior is not captured within conventional linear-response theory or purely unitary scattering models.\\
\textit{Combined unitary–projective evolution (right):} When symmetry-broken Hamiltonian propagation is combined with localized projections, qualitatively new functionalities can emerge, including directional transmission probabilities (single-particle valve behavior), nonreciprocal transport of coherence, and, at finite temperatures $T$, non-equilibrium steady-state circulating currents $I$ generating magnetic moments $M$. Directionality arises from the asymmetric exposure of counter-propagating trajectories to projective updates.}
\label{fig:Fig_E}
\end{figure}

Against this backdrop, one might ask whether materials or devices in general can be deliberately engineered to benefit from the combination of unitary and projective dynamics. Indeed, purposefully combining engineered coherent evolution with tailored projection processes establishes a unifying design principle for advanced materials and devices \cite{Mannhart2018, Bredol2021b, Mannhart2021}. Figure~\ref{fig:Fig_E} summarizes the core elements of this principle. In such unitary--projective (u-p) materials and devices, projection processes do not merely assist transport and stabilize phases, but also fundamentally reshape the admissible steady states and response properties. In practice, this design strategy can be formulated, for example, within the quantum-trajectory framework, where individual electronic states propagate unitarily between stochastic projection events (Fig.\,\ref{fig:Fig_B}(B)). Material properties then emerge from ensemble averages over such trajectories.

With this conceptual structure in place, we will now discuss how u-p dynamics can be realized in concrete material architectures.

\FloatBarrier

\section{Quantum-Material Design Based on Combined Unitary–Projective Evolutions}

\subsection{Designing State Projections into Electron Systems of Materials}

Whereas incorporating Schr\"odinger dynamics into materials modeling is standard practice, the targeted inclusion of projection or measurement processes into materials may at first appear less straightforward. It is important to emphasize, however, that such projections do not require a conscious observer or a conventional measurement apparatus. In the decoherence model, for instance, the mere leakage of information into environmental degrees of freedom is sufficient to induce decoherence. Tracing out the environment renders this process equivalent to a state projection in the reduced system \cite{ Breuer2002, vonNeumann1932, Zurek2003}. Physical collapse models, although conceptually distinct (see, e.g. \cite{Bassi2013}), effectively lead to the same practical consequences for materials design.

To deliberately insert projection processes that enable unitary–projective functionality into a material’s electronic system, two conditions must be satisfied: First, information must be able to leak into the environment. Second, the system’s dynamics must include projection events that induce irreversible, stochastic population transfer between symmetry-defined sectors of the system’s Hilbert space. Both requirements can be easily met in practical material architectures. For instance, a channel for information leakage can be created by introducing local inelastic electron-phonon scattering centers and allowing the phonons to dissipate into a thermal bath.

Nontrivial sequences of such combined evolutions can be engineered in many ways. As a simple illustrative example, consider a semiconductor designed such that an electron wave packet propagates ballistically (unitarily) in the conduction band. Upon encountering an in-gap trap state, the electron may become localized at the defect after emitting a phonon into the bath, corresponding to a projection event. The electron then remains trapped until a subsequent phonon-mediated process releases it back into a delocalized conduction-band state, constituting a second projection event.

\subsection{Incorporating Function-Enabling Unitary Dynamics into Such an Electron System}

The trapping and detrapping dynamics described above occur ubiquitously in solids. In conventional situations, however, unitary evolution treats forward and backward propagation equivalently, i.e., propagation from the left contact to the right contact and vice versa. In these settings, projection steps merely interrupt ballistic propagation, manifesting as ordinary noise or, under non-equilibrium conditions, as dissipation.

\begin{figure}[t]
\centering
\captionsetup{width=0.9\textwidth}
\includegraphics[width=0.9\textwidth]{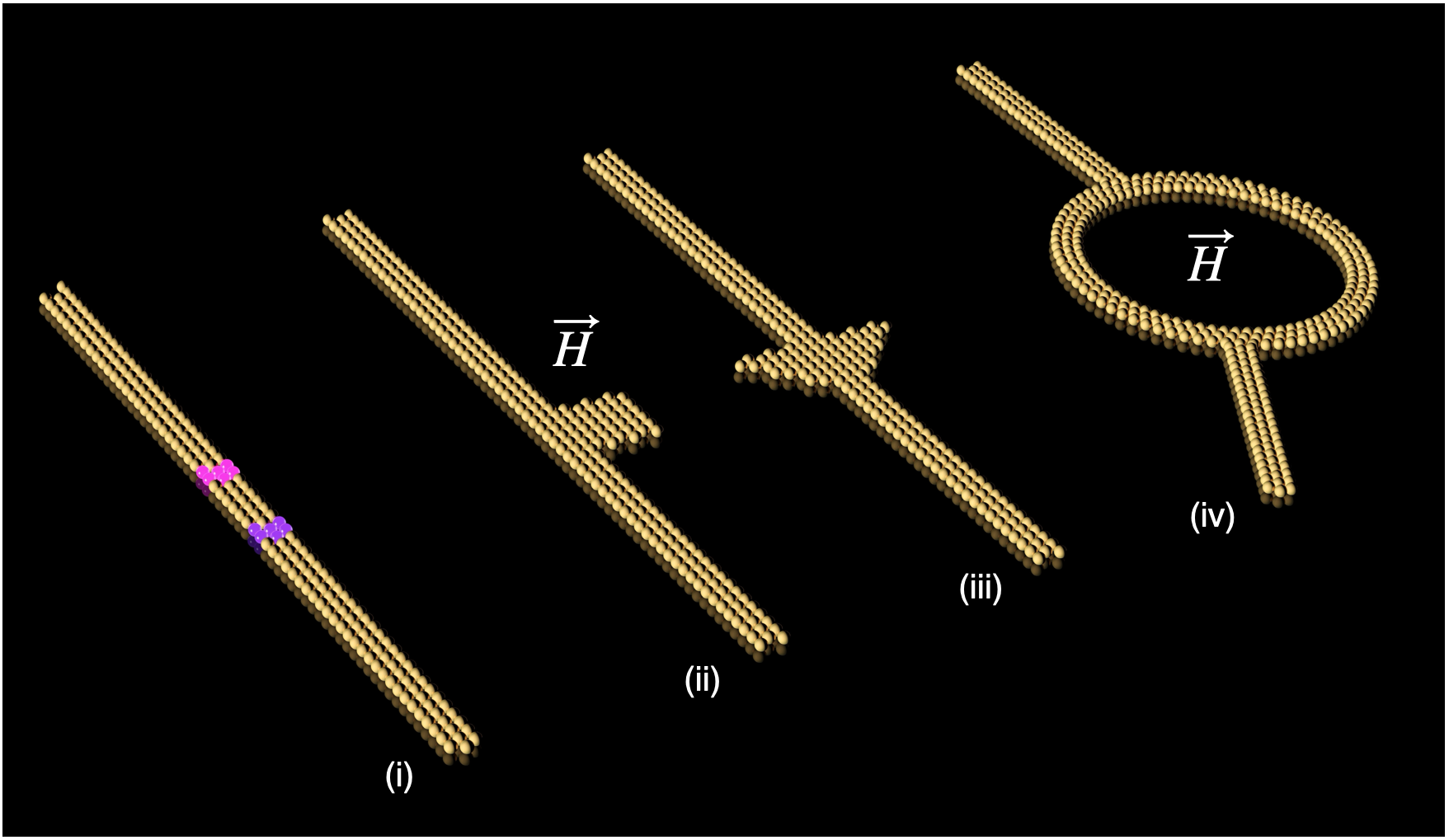}
\vspace{3 mm}
\caption{\textbf{Asymmetric nanoscale conductors.} From left to right, the figure shows (i) a nominally symmetric one‑dimensional conductor containing two tunnel barriers of different heights (pink and purple, respectively), (ii) a conductor with transverse asymmetry, (iii) a conductor with longitudinal asymmetry, and (iv) an asymmetric Aharonov–Bohm ring. To generate directional asymmetry in ballistic electron propagation along the transport direction, the transversely asymmetric conductor (ii) and the quantum ring (iv) must be subjected to a perpendicular magnetic field $H$. In contrast, conductors (i) and (iii) possess the necessary spatial asymmetry without a magnetic field. Figure adapted from \cite{Bredol2021b}.}
\label{fig:Fig_F}
\end{figure}

Moving beyond these standard situations, the freedom to design the architecture of molecules, unit cells, or nanostructures makes it possible to engineer a unitary propagation that already breaks spatial symmetry, for example, by implementing asymmetric geometries or by introducing electronic properties that are not spatially symmetric. Such symmetry breaking can be achieved, for instance, by arranging sequences of ions with non-identical on-site potentials \cite{Bredol2021b} or by locally employing Rashba spin–orbit coupling within an otherwise spatially symmetric structure \cite{Mannhart2018}. Even in homogeneous materials, suitably chosen spatially asymmetric geometries that control electron-wave interference can achieve the same effect, as illustrated by the examples in Fig.\,\ref{fig:Fig_F}. Note that a perpendicular magnetic field must be applied when only the symmetry perpendicular to the direction of electron transport is broken \cite{Bredol2021a}.

\FloatBarrier

The transmission probabilities remain reciprocal in all of these structures, yet the time an electron spends inside a structure can depend on the direction of its overall motion. In mathematical terms, unitarity and time-reversal symmetry fix the transmission probabilities but still leave sufficient freedom in the internal phase structure of the scattering matrix for the Wigner–Smith delay times \cite{Wigner1955, Smith1960} to be nonreciprocal. In this way, one obtains a unitary evolution in which forward and backward propagation exhibit different temporal profiles \cite{Bredol2021a}. 
Even so, the evolution must preserve the balance of occupations between equally populated, symmetry-related transport channels. This balance is enforced by the unitarity of the scattering matrix.

It therefore becomes clear that when symmetry-broken unitary evolution is combined with nonunitary projection steps that stochastically transfer electrons between symmetry-related channels, qualitatively new transport properties can emerge, particularly when the effective switching rates depend on the direction of the electron propagation. To illustrate this, we now present two examples.

\subsection{Examples of Systems with Combined Unitary–Projective State Evolutions Achieving Novel Functionalities}

We begin by discussing systems in which the combination of unitary and projective state evolution yields two functionalities that lie outside the scope of standard device physics, namely nonreciprocal single-electron transmission in a passive system and a novel category of magnetism driven by thermal energy. These functionalities arise from electron trapping and detrapping events that stochastically interrupt a symmetry-broken unitary evolution. Because these effects rely on geometrical asymmetries and open‑system, projection‑induced dynamics, they are not captured by the foundational band diagram framework. 

We also note that the model calculations that can be practically carried out for such systems necessarily rely on simplifying assumptions, are not exact, and may inherit limitations from the underlying theoretical formalisms. In the Lindblad calculations presented here, for instance, the projection operators are phenomenological rather than rigorously derived from a fully microscopic treatment of the traps and their environment. Moreover, the formalism relies on standard approximations, such as the Markovian and secular limits. It also assumes weak system–environment coupling and is limited to single-particle dynamics. Accordingly, the present models should be understood as descriptions aimed at identifying causal physical mechanisms and robust qualitative behavior, rather than as definitive quantitative predictions for specific material systems.

\subsubsection{Example 1: Passive Single-Electron Valves}

To introduce the basic operating principle of these devices, we start from a conventional Aharonov–Bohm ring \cite{Aharonov1959}, introduce a controlled spatial asymmetry, and analyze electron transport in terms of propagating wave packets rather than retreating to stationary energy eigenstates. Such symmetry-broken quantum rings can be realized in asymmetric organic molecules or engineered nanostructures, for example.

\begin{figure}[tbp]
\centering
\captionsetup{width=0.9\textwidth}
\includegraphics[width=0.9\textwidth]{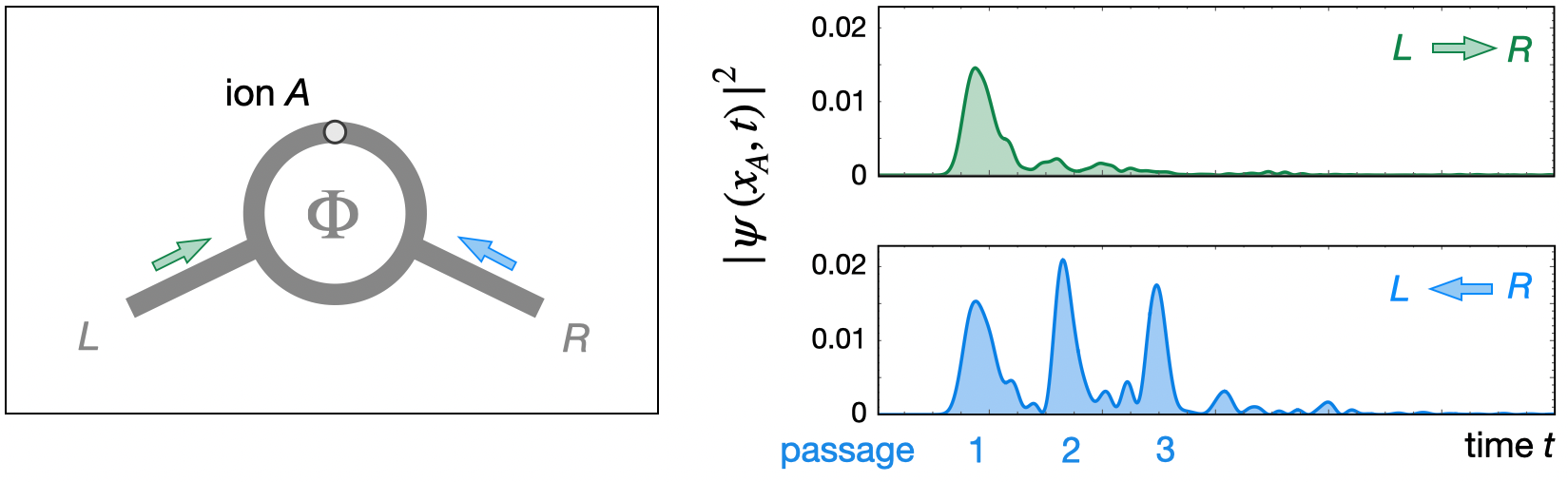}
\caption{\textbf{Ballistic transport in an asymmetric quantum ring.} Time evolution of unitary electron propagation through the ring from the left contact $L$ to the right contact $R$ (green) and from $R$ to $L$ (blue). The schematic on the left depicts the ring, threaded by an applied magnetic flux $\Phi$, with an ion $A$ (white circle) embedded at the center of its upper arm. The panels on the right show quantitative results for the time-dependent probability density of the electron wave function at the position of ion $A$, obtained from tight-binding calculations \cite{Bredol2021b}. For propagation from $L$ to $R$ (top panel), the ion is encountered once, whereas for propagation from $R$ to $L$ (bottom panel), the electron passes the ion three times. The oscillatory “ringing” in the probability density  predominantly arises from the finite number of tight-binding sites. Figure adapted from \cite{Bredol2021b}.}
\label{fig:Fig_C}
\end{figure}

Asymmetric quantum rings, as illustrated in the left panel of Fig.,\ref{fig:Fig_C}, impose two distinct phase shifts on wave packets that propagate along their two branches. One phase shift is geometric, arising from the difference in path length between the two arms. The other phase shift is generated by a possible magnetic flux $\phi$ threading the ring. Because only the latter—the Aharonov–Bohm phase shift—depends on the direction of wave-packet circulation around the ring, suitable choices of the path-length difference and magnetic flux give rise to a remarkable interference behavior between the two partial wave packets into which an electron splits when entering the ring.

For an applied flux $\phi = (n + 1/4) \; \Phi_0$ , where $n$ is an integer, $\Phi_0 = h/e$ the flux quantum (with $h$ Planck’s constant and $e$ the elementary charge), and for a path-length difference $\Delta l = {\pi} / {2k}$, with $k$ the wave number at the center of the wave packet, the two partial wave packets propagating from left to right acquire a relative phase shift that is an even multiple of $\pi$. This results in a straightforward electron trajectory: an electron incident from the left contact splits into two partial wave packets that traverse the two arms of the ring, recombine constructively at the right junction, and exit the device towards the right contact.

The situation is qualitatively different for an electron propagating from right to left. For the same applied flux, the two partial wave packets into which the electron splits acquire a relative phase shift equal to an odd multiple of $\pi$. As a result, the trajectory becomes multiply reflected. An electron incident from the right contact again splits into two partial wave packets that propagate along the two arms of the ring. When these packets reach the left junction, they have a relative phase difference of an odd multiple of $\pi$, which leads to destructive interference. This interference prevents the packets from recombining and exiting the device. Instead, they are fully back-reflected. On the return path, the wave packets accumulate an additional phase difference of an even multiple of $\pi$ because they propagate from left to right. Upon reaching the right junction they interfere destructively once more and are reflected. During their second traversal from right to left, the two wave packets gain an additional odd-$\pi$ phase shift. They then arrive at the left junction with an overall phase difference equal to an even multiple of $\pi$, interfere constructively, and finally exit the system towards the left contact.

Consequently, the time an electron spends inside the ring differs by a factor of about three, depending on whether it travels from the left contact to the right contact or vice versa. Moreover, any atom or ion in the ring is exposed to a passing electron wave packet once when an electron travels from the left contact to the right, but three times when an electron propagates from the right contact to the left. This behavior is illustrated in Fig.\,\ref{fig:Fig_C}, which shows the result of a corresponding tight-binding calculation of the Schr\"odinger equation. These tight\--binding results are also provided as videos in the Supplementary Material.

\begin{figure}[tbp]
\centering
\captionsetup{width=0.9\textwidth}

% -------- Panel (A) --------
\begin{subfigure}{\textwidth}
  \raggedright
  \textbf{(A)}\\[0.6em]
  \centering
  \includegraphics[width=0.6\textwidth]{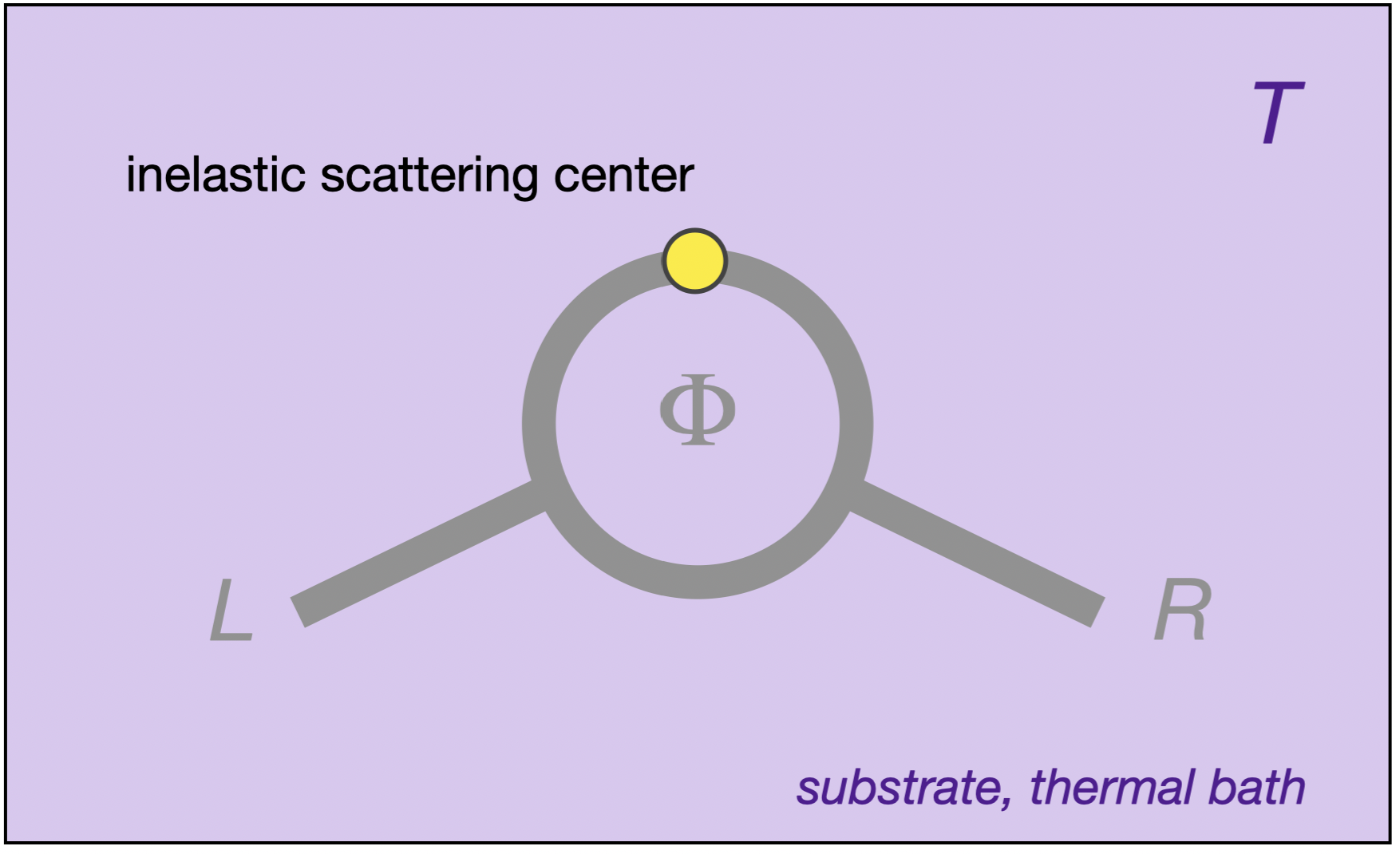}
\end{subfigure}

\vspace{4.2em}

% -------- Panel (B) --------
\begin{subfigure}{\textwidth}
  \raggedright
  \textbf{(B)}\\[0.6em]
  \centering
  \includegraphics[width=0.8\textwidth]{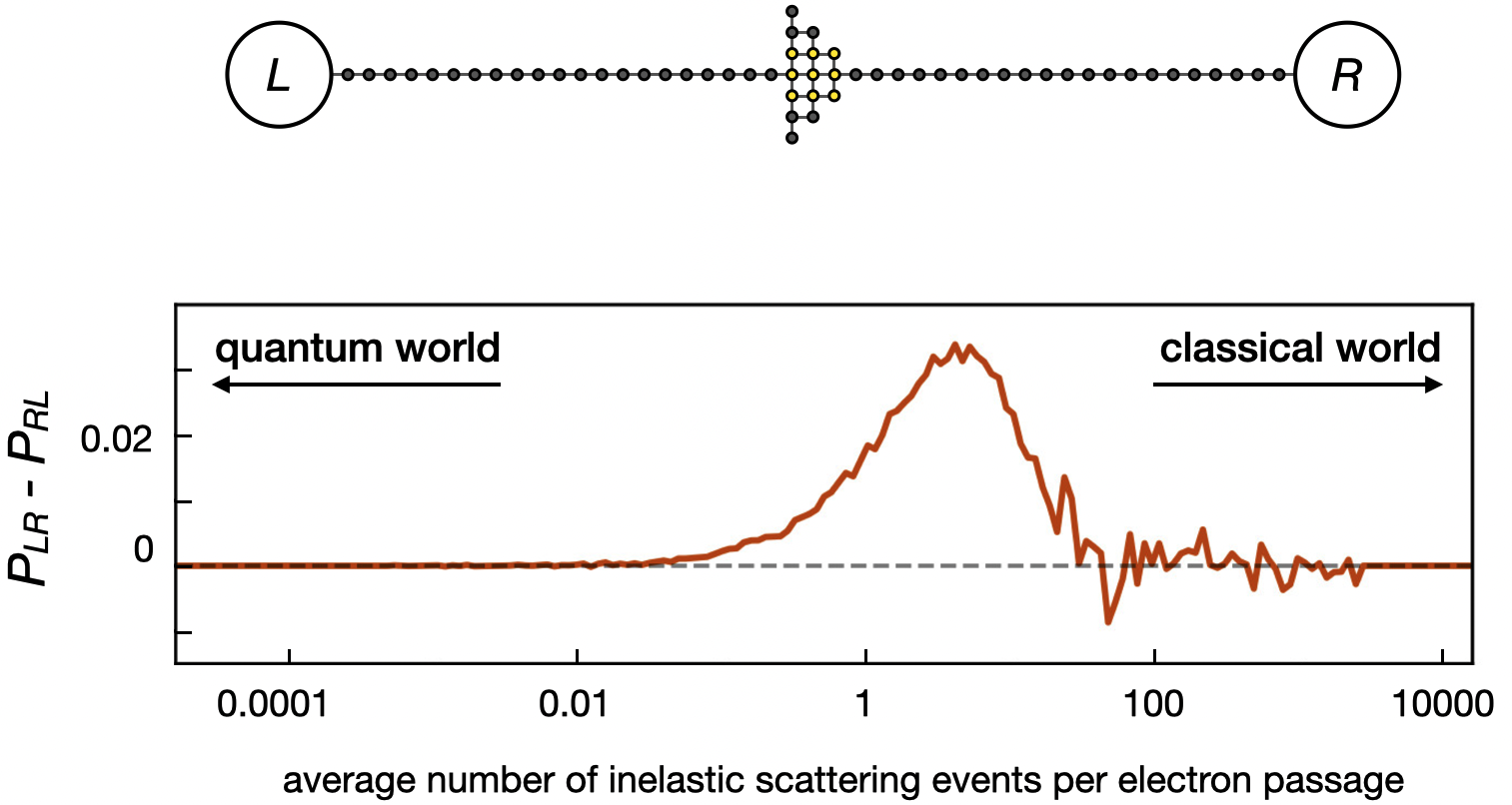}
\end{subfigure}

\vspace{1.0em}

\caption{\textbf{Unitary–projective transport in asymmetric structures.}\\
(A) Schematic of an asymmetric quantum ring connecting contacts $L$ and $R$, threaded by a magnetic flux $\Phi$. The ring contains an inelastic scattering center (yellow dot), such as an electron trap, at the center of its upper arm. This center couples the electronic degrees of freedom of the ring to the phonon system of the environment (e.g., a substrate, lila) at temperature $T$, which may or may not be thermally coupled to a further external bath (e.g., the atmosphere).\\
(B) Difference between the transmission probabilities $P_{L\to R}$ and $P_{R\to L}$ calculated as a function of the average number of inelastic scattering events per electron passage. The calculations are based on quantum-trajectory simulations \cite{Bredol2021b} for the structure sketched at the top, where the dots represent the sites of the tight-binding model. Yellow sites mark electron traps. Nonreciprocal transmission arises for an intermediate range of inelastic scattering, approximately 0.05–50 scattering events per passage. In the purely unitary quantum limit with negligible scattering and in the classical limit with a large number of scattering events leading to diffusive transport, the transmission is reciprocal, in accordance with Onsager’s reciprocity theorem. Figure adapted from \cite{Bredol2021b}.}
\label{fig:Fig_RP}
\end{figure}

Having defined the unitary part of the electron propagation, we now turn to the interaction of an electron wave packet with a trap state embedded in such a ring (Fig.\,\ref{fig:Fig_RP}(A)). Upon encountering the trap, there is a probability $p$ that the electron will be captured and subsequently re-emitted into the conduction band. We model this process as Markovian; once an electron is trapped and a phonon is emitted, the electronic subsystem retains no information about the electron's original direction. 
Consistent with this Markovian framework, the detrapping phonons are treated as incoherent wave packets of finite spatial extent, set by phonon lifetimes and scattering processes. The propagation directions of these wave packets are drawn from a thermal distribution. Assuming the trap is coupled symmetrically to left- and right-propagating electronic states, the electron, once detrapped, continues toward the left or right contact with equal probability. Consequently, each encounter with the trap is characterized by a backscattering probability of $p/2$.

It is again emphasized that, within the standard open-system description based on globally unitary system–environment dynamics, information is, of course, neither created nor destroyed. Rather, during trapping events, information is transferred from the electronic subsystem to environmental degrees of freedom such as phonons, while during detrapping, a corresponding amount of new information is injected from the environment back into the electronic subsystem. Therefore, there is no net information loss or gain in the combined system, despite the electronic subsystem undergoing repeated non-equilibrium exchanges of information with its environment. Importantly, the device functionality relies on this cyclic process, wherein information about prior trajectory occupations is replaced by stochastic environmental input which in turn determines the occupation of subsequent transport pathways. At the level of the reduced electronic dynamics, this replacement corresponds to stochastic population transfer between symmetry-defined sectors of the Hilbert space associated with the unitary evolution (e.g., left-moving versus right-moving transport channels). Because the underlying unitary dynamics is asymmetric, the rates of these projection-induced transfers depend on the sector in which the state initially resides, enabling directional transport and non-equilibrium steady states.

\FloatBarrier

According to these considerations, introducing a single trap state into a quantum ring tuned to the 1:3 dwell-time condition makes the transport direction dependent.
Electrons injected from the left and propagating towards the right contact encounter the trap effectively only once. At that encounter, the probability that the electron is captured and subsequently re-emitted back towards the left contact is $P^{L\to R}_{\mathrm{back}} \; = \; p/2$.

Conversely, under the unitary Hamiltonian evolution, electrons injected from the right zigzag through the ring three times before exiting to the left. In this simple probabilistic picture, this implies that such electrons encounter the trap three times before finally leaving the structure. The corresponding backscattering probability is therefore $P^{R\to L}_{\mathrm{back}}\; = \; (1+ (1-p) + (1-p)^2) \; p/2$, where the three terms denote the probabilities of backscattering at the first, second, and third encounters with the trap, respectively. For small $p$, this sum reduces to $P^{R\to L}_{\mathrm{back}} \approx 3\,\frac{p}{2}$. The overall transmission probability of an electron is therefore direction dependent.

This directionality is remarkable because it is strictly prohibited in standard materials and conventional two-terminal devices. In the present case, however, it naturally arises from the presence of an inelastic trap that induces stochastic projection events, thereby converting direction-dependent dwell times of the underlying unitary evolution into genuinely nonreciprocal transport.

The above discussion was intentionally heuristic. However, as shown in \cite{Bredol2021b}, an explicit quantum-mechanical treatment of the trap and its environment likewise yields this directional asymmetry in the transmission probability. 
In that study, tight-binding calculations were performed for asymmetric systems with even greater structural simplicity than the asymmetric ring, namely singly connected, arrow-shaped devices (Fig.\,\ref{fig:Fig_RP}(B), top). In such structures, unitary propagation again leads to direction-dependent dwell times of electrons inside the device (see Supplementary Material for videos), resulting in an asymmetry in the probability of encountering and scattering from embedded defects. To elucidate how this direction-dependent exposure to defects influences transport, the calculations include trap states located in the central region of the structure.

The transmission through such a structure is evaluated using a quantum-trajectory formalism or, with consistent results, an equivalent Lindblad master-equation approach. The results of these calculations are summarized in Fig.\,\ref{fig:Fig_RP}(B), which displays the resulting nonreciprocity of the transmission probability, $ P_{\mathrm{sort}} = P_{L \to R} - P_{R \to L}$, for such a device.

The nonreciprocity, $P_{\mathrm{sort}}$, is evaluated as a function of the scattering probability $p$, which again denotes the probability that an electron will interact with the trap during a single passage.
These calculations reveal that the devices indeed exhibit directional transmission, as evidenced by a pronounced peak in $P_{\mathrm{sort}}$ as a function of $p$. The maximum directionality of $\approx 2-3 \%$, occurring for five to ten scattering events per electron passage, is fully consistent with the intuitive picture developed above.

Related behavior has been demonstrated by Ferreira \textit{et al.} \cite{Ferreira2024}, who derived exact non-equilibrium Green’s function expressions within the Keldysh formalism for transport in continuously monitored, noninteracting fermionic systems, in the spirit of Meir--Wingreen theory. They show that monitoring induces an inelastic current component that depends nontrivially on the monitoring strength $\gamma$, and that these monitor-induced inelastic processes can generate nonreciprocal currents. This includes finite currents at zero bias under appropriate symmetry conditions. For the specific case of monitoring a single-level occupation and averaging over measurement outcomes, this is formally equivalent to coupling the system to an infinite-temperature bosonic bath. The central mechanism, measurement-induced nonunitary reduced dynamics leading to nonreciprocal transport, closely parallels the projection-induced directionality discussed here, although their setup does not involve trap sites. In particular, the induced current exhibits a pronounced maximum as a function of $\gamma$, reflecting the same competition between coherent dynamics and stochastic interruptions that underlies the maximum in nonreciprocal transmission versus scattering probability shown in Fig.\,\ref{fig:Fig_RP}(B), and the temperature-driven maximum in nonreciprocity discussed in Sect.~\ref{subsec:T_peak}.

\begin{figure}[tbp]
\vspace*{-0.5cm}
\centering
\captionsetup{width=0.9\textwidth}
\includegraphics[width=0.9\textwidth]{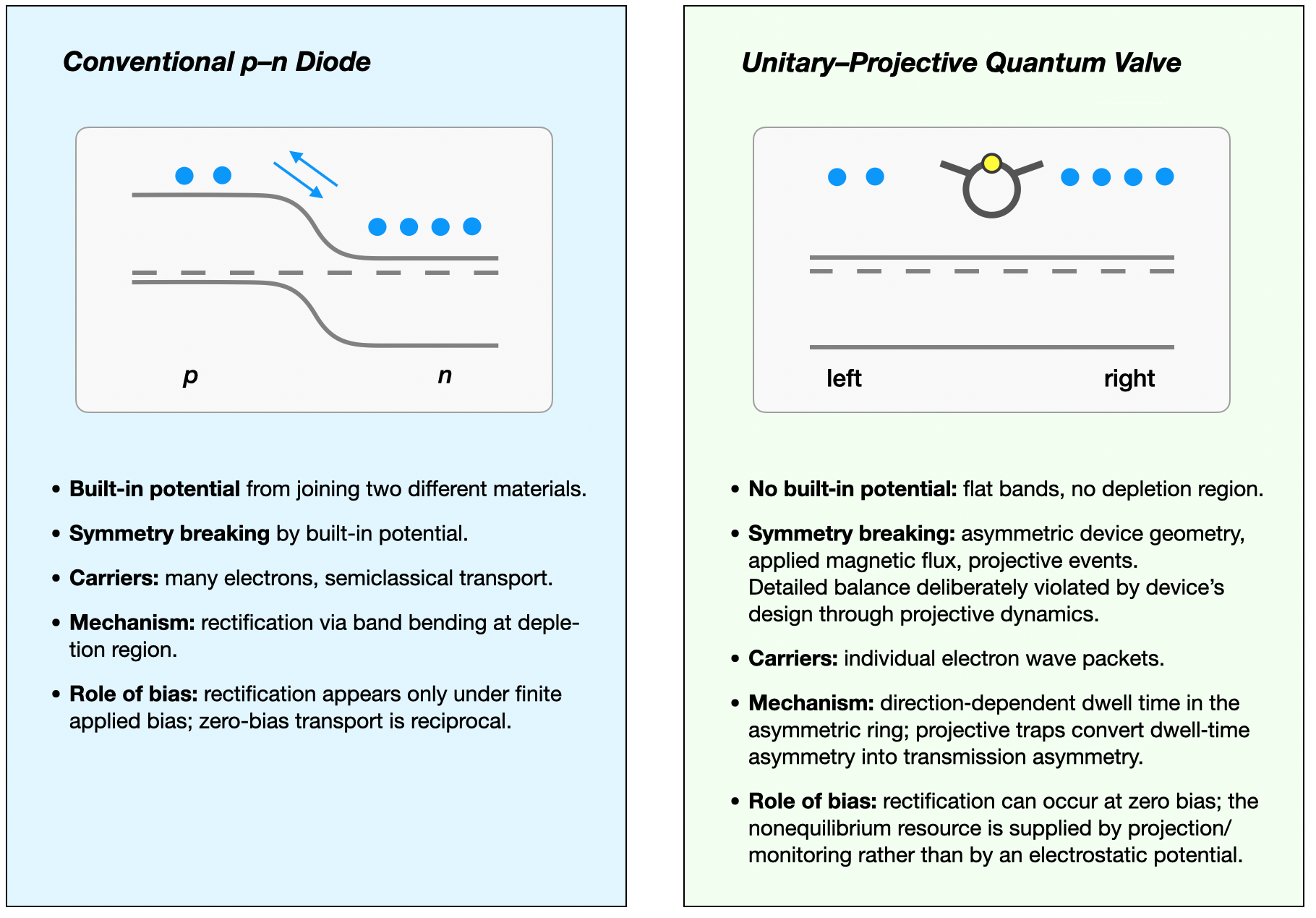}
\vspace{3 mm}
\caption{\textbf{Comparison of rectification mechanisms in conventional p–n diodes and unitary–projective valves.}\\
Left panel: In a conventional p–n diode, rectification arises from the static built-in electrostatic potential and the associated depletion region (band bending). Rectification requires a finite voltage or current bias and results from semiclassical many-carrier drift–diffusion transport.\\
Right panel: In a u-p diode, rectification emerges from a dynamical asymmetry involving direction-dependent dwell times and stochastic projection (measurement) events. These events render the reduced dynamics effectively irreversible and can evade detailed balance. This mechanism operates on individual electron wave packets and can generate rectification without a static energy barrier in regimes where conventional reciprocal transport constraints need not apply; the non-equilibrium bias is supplied by the projection/monitoring process rather than by band bending. Note that the projection‑driven device mechanism is not captured by the band diagram.
}
\label{fig:Fig_D}
\end{figure}

It is important to emphasize that this valve behavior differs fundamentally, in both mechanism and characteristic properties, from that of conventional diodes such as p–n junctions. Conventional diodes rely on material interfaces. These interfaces produce rectification only in the nonlinear response regime. In contrast, the u-p valve function discussed here arises from symmetry-broken dynamics and projection-induced state updates. These distinctions are summarized in Fig.\,\ref{fig:Fig_D}.

%\FloatBarrier

We emphasize that any directionality $P_{\mathrm{sort}} > 0$ lies outside the predictions of textbook transport physics. This directional transmission is inaccessible in standard reciprocal two-terminal systems governed by the reciprocity and linear-response constraints discussed in Sect.~\ref{sec:Evolution_of_States}. Consistent with this picture, the calculated directionality shown in Fig.\,\ref{fig:Fig_RP}(B) vanishes for $p < 10^{-2}$ (purely unitary quantum limit) and for $p > 10^{2}$ (classical diffusive regime).

These results imply that such devices act as passive valves for single particles \cite{Bredol2021b}. This behavior may have far‑reaching consequences. For example, one could envision quantum materials that, even in uniform bulk form, rectify individual carriers, or devices that operate as diodes at extremely small currents and voltages. From a device-engineering perspective, the absence of material interfaces is of interest because it eliminates interface-based effects, such as depletion layers, which typically constrain capacitance, resistance, and operating frequencies. Beyond transport rectification, non-reciprocity also manifests in the coherence properties of the propagating wave packets. Wave packets moving along the low-transmission direction undergo enhanced decoherence due to repeated projection events. In contrast, those traveling in the opposite direction retain a higher degree of coherence. The effective inductance of the system likewise exhibits corresponding non-reciprocity.

This novel diode functionality has, to some extent, already been observed experimentally in a recently developed microwave-based model system \cite{Chen2024}. In that realization, the unitary propagation of electron wave packets is emulated by propagating microwave pulses, an asymmetric quantum-ring geometry is implemented with standard microwave circuitry, and dephasing or projection are represented by controllable dissipative attenuation. As reported in \cite{Chen2024}, the device exhibits asymmetric transmission for coherent pulses and even for incoherent broadband noise, closely resembling the behavior predicted for quantum rings containing dephasing centers. Notably, the measurements also reveal the characteristic 3:1 ratio in the direction-dependent dwell times. The experimental confirmation of this ratio strongly supports the idea that broken spatial symmetry combined with stochastic interruptions of coherent propagation naturally leads to directional, nonreciprocal transport.

\begin{figure}[tbp]
%\vspace*{-0.0cm}
\centering
\captionsetup{width=0.9\textwidth}
\includegraphics[width=0.9\textwidth]{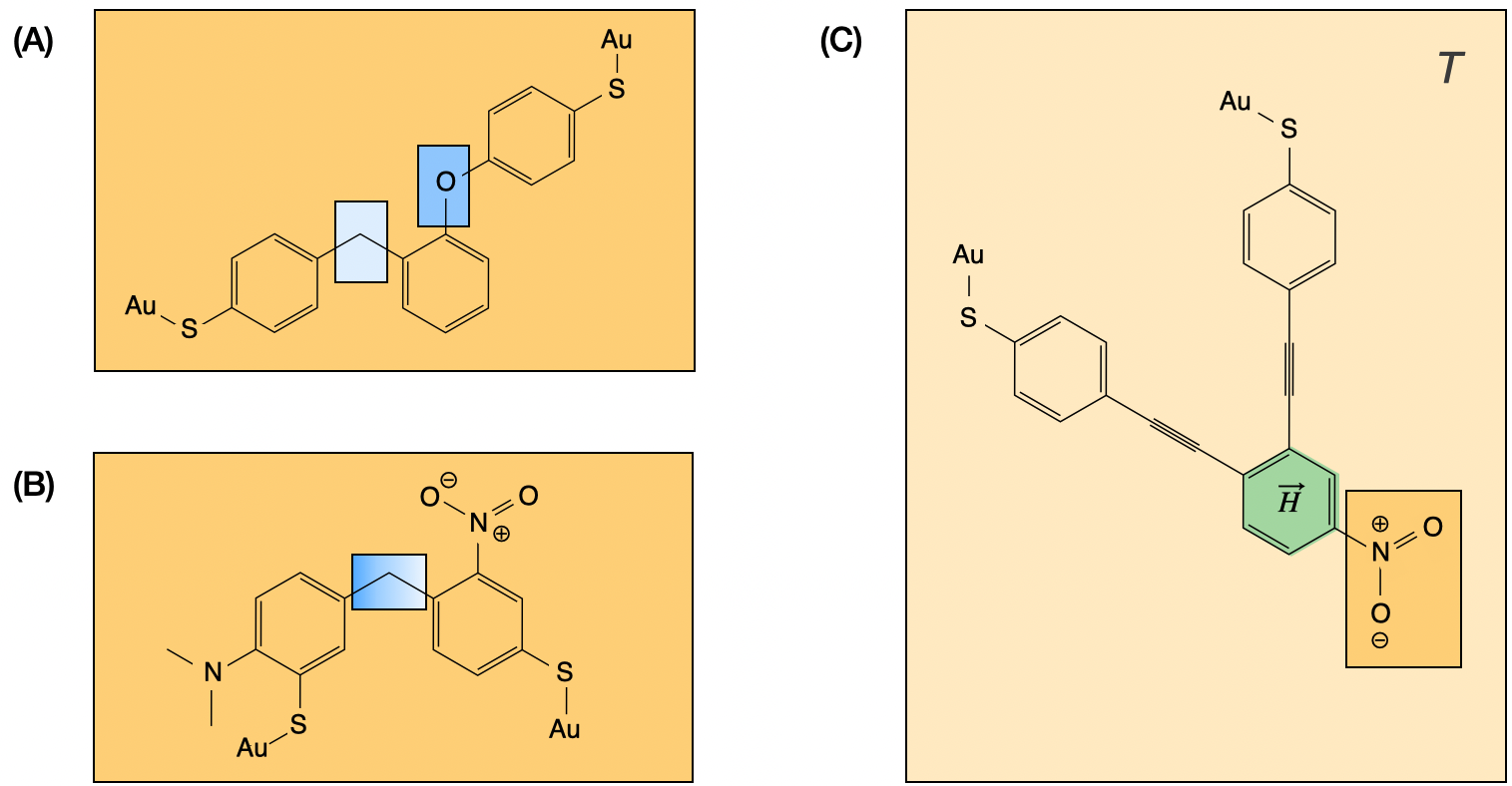}
\vspace{3 mm}
\caption{\textbf{Conceptual embedding of unitary–projective transport with nonreciprocal electron transmission using single organic molecule conductors.}\\
\textbf{(A)} Phenyl–$\textrm{CH}_2$–phenyl–O–phenyl molecular wire, in which the conducting $\pi$-conjugated backbone is interrupted by two chemically distinct tunnel barriers: a methylene linker (–$\textrm{CH}_2$–, light blue) and an ether linker (–O–, dark blue) which connect phenylene units. The ether linker constitutes a particularly high tunnel barrier for electron transport.\\
\textbf{(B)} Donor–$\sigma$–acceptor molecular wire, dimethylamino–phenyl–$\textrm{CH}_2$–phenyl–nitro, in which a single methylene (–$\textrm{CH}_2$–) segment connects electronically inequivalent donor and acceptor rings, resulting in an intrinsically asymmetric (tilted) tunnel barrier (blue shaded region).\\
\textbf{(C)} Oligo(phenylene–ethynylene) molecular wire, 4-nitro-1,3-bis(ethynylphenyl)benzene, containing an intrinsically asymmetric aromatic ring. A magnetic field $H$ applied perpendicular to the ring plane (green) and the molecular asymmetry break time-reversal symmetry and reciprocity. The nitro group on the central ring provides a localized, strongly coupled vibrational manifold and a polar site that enhances inelastic electron–vibration interactions.\\
In all cases, coupling to the substrate or a thermal bath occurs via inelastic scattering into the substrate phonon bath (symbolized by the orange color). These three structures illustrate generic molecular design motifs for implementing unitary–projective transport, rather than optimized candidates for specific device functions.}
\label{fig:Fig_M}
\end{figure}

\begin{figure}[tbp]
\centering
\renewcommand\thesubfigure{\Alph{subfigure}}

\begin{subfigure}[t]{\textwidth}
  \centering
  \includegraphics[width=0.95\textwidth]{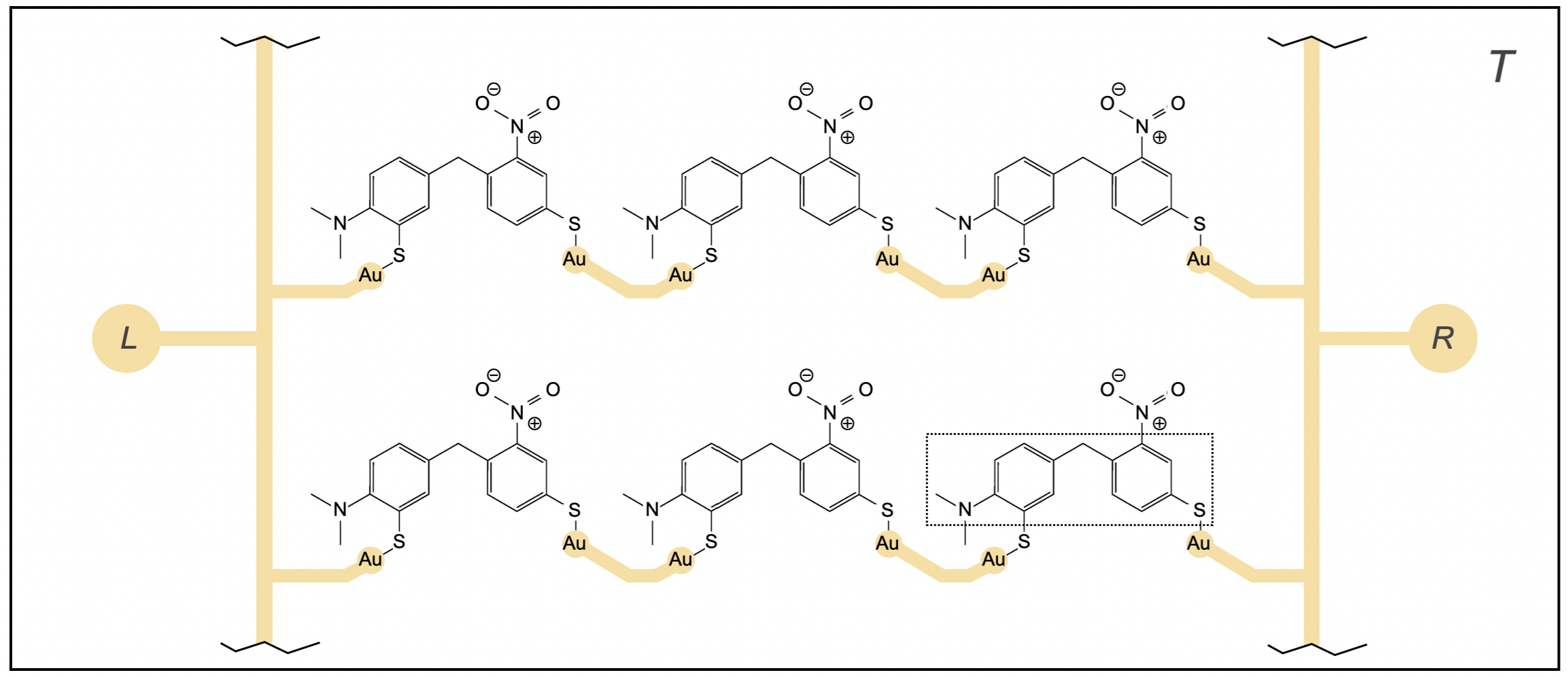}
  \caption{} % leave empty if you want only (A) without text
  \label{fig:Mac-a}
\end{subfigure}\hfill
\vspace{3em} % adjust vertical spacing between panels
\begin{subfigure}[t]{\textwidth}
  \centering
  \includegraphics[width=0.8\textwidth]{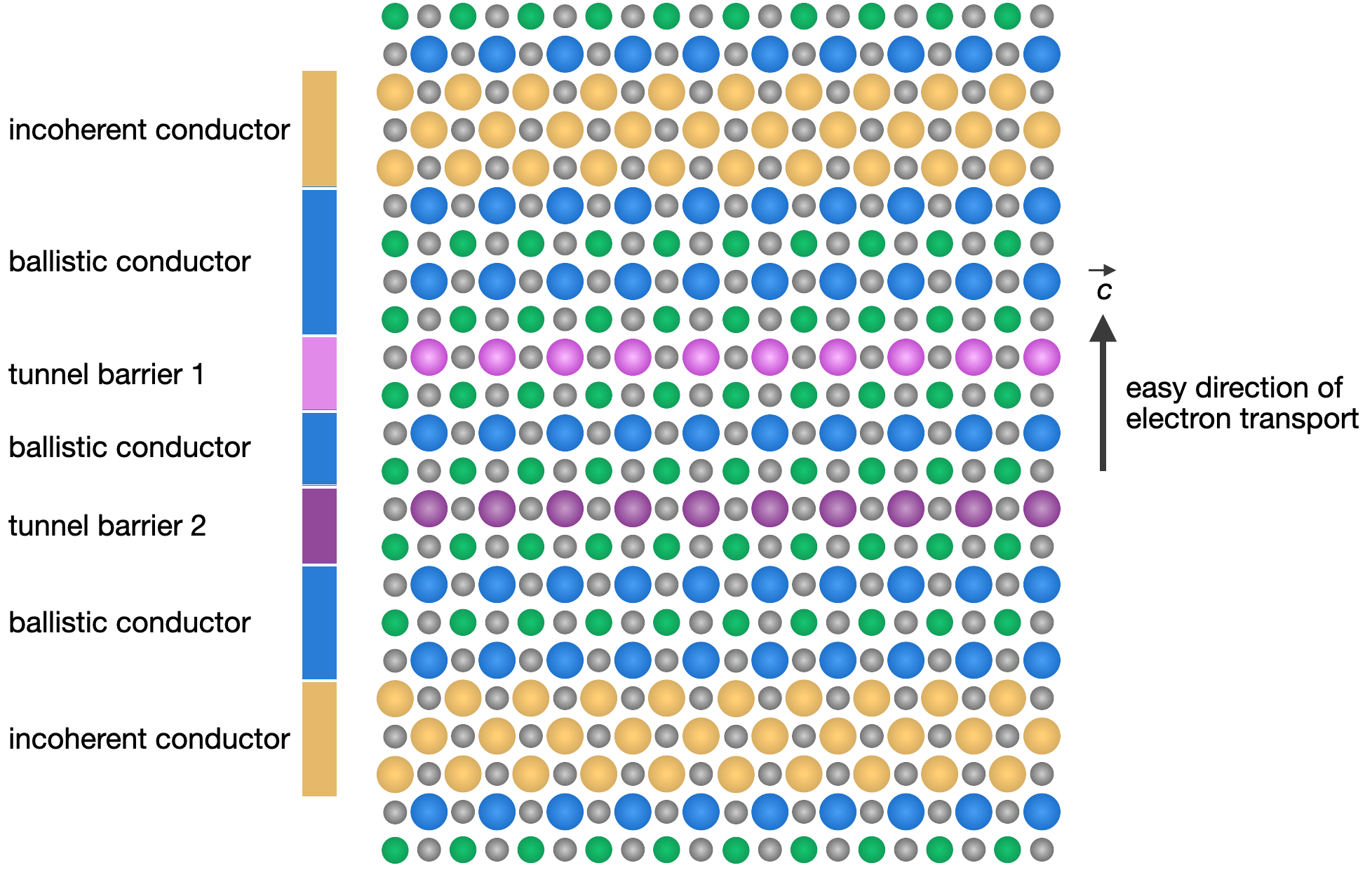}
  \caption{}
  \label{fig:Mac-b}
\end{subfigure}

\vspace{0.4em}
\caption{\textbf{Schematic principles for scaling unitary–projective devices to macroscopic dimensions.}\\
\textbf{(A)} Macroscopic implementation of u-p materials based on molecular assemblies. An array of donor–$\sigma$–acceptor molecular wires (cf. Fig.\,\ref{fig:Fig_M}(B)) is connected to metallic contacts (gold), which act as incoherent electron reservoirs. The requirement that an electron undergo on the order of one inelastic scattering event per passage applies only to individual molecular units, as exemplarily indicated by the framed molecule in the lower right.\\
\textbf{(B)} Macroscopic implementation based on layered crystalline structures. A cross section of a possible three-dimensional realization of the structure in Fig.\,\ref{fig:Fig_F}(i) is shown. The structure consists of two tunnel barriers of different heights (pink and purple) embedded in a quasiballistic conductor (blue). The gold-colored layers denote incoherent conductors. The structure represents part of a crystalline superlattice with electron transport along the crystallographic \(c\)-axis.}

\label{fig:Fig_Mac}
\end{figure}

At first sight, creating an analogous system in a solid state seems simple. For instance, one could use nanopatterned devices made from materials with sufficiently large mean free paths. In these systems, the projection steps could naturally occur through inelastic electron-phonon scattering. 
However, translating this conceptual simplicity into a practical device presents significant challenges. In particular, achieving the necessary patterning accuracy is difficult; the device geometry must be defined with a precision far greater than the Fermi wavelength. Even in GaAs heterostructures, the Fermi wavelength is only $\lambda_{\mathrm{F}} \approx 40-80\,\mathrm{nm}$. To circumvent these stringent fabrication requirements, one could consider device platforms in which ballistic transport with broken symmetry arises from a macroscopic quantum effect, such as a quantum Hall effect. In this case, the projection steps could be implemented by appropriately engineered inhomogeneities. Alternatively, one may use asymmetric molecular or crystalline structures, with inelastic processes providing the necessary projection events (Fig.\,\ref{fig:Fig_M}, Fig.\,\ref{fig:Fig_Mac}).

The considerations presented above assume that electrons are driven by standard voltage or current sources. We now turn to unitary–projective dynamics initiated under thermal equilibrium conditions.

In thermal equilibrium, electron motion is not induced by non-equilibrium distributions imposed by external sources. Instead, it arises, for example, from Johnson–Nyquist noise \cite{Johnson1928, Nyquist1928}, which corresponds to stochastic thermal fluctuations producing a continuous stream of electrons and electron–hole excitations entering and leaving an electronic system, even in the absence of an applied bias. This raises a central question: How do unitary–projective materials respond when driven not by externally imposed non-equilibrium distributions, but instead by intrinsic thermal fluctuations?

%\clearpage

\subsubsection{Example 2: Novel Category of Magnetism}
\label{subsection:novel_magnetism}

In this second example, we explore whether electron wave packets generated by Johnson–Nyquist noise can also generate functionalities inaccessible within a purely Hamiltonian framework. By doing so, we show that u-p evolutions can produce a novel category of magnetism.

\begin{figure}[tbp]
\vspace*{-1.0cm}
\centering
\renewcommand\thesubfigure{\Alph{subfigure}}

\begin{subfigure}[t]{0.48\textwidth}
  \centering
  \includegraphics[width=\textwidth]{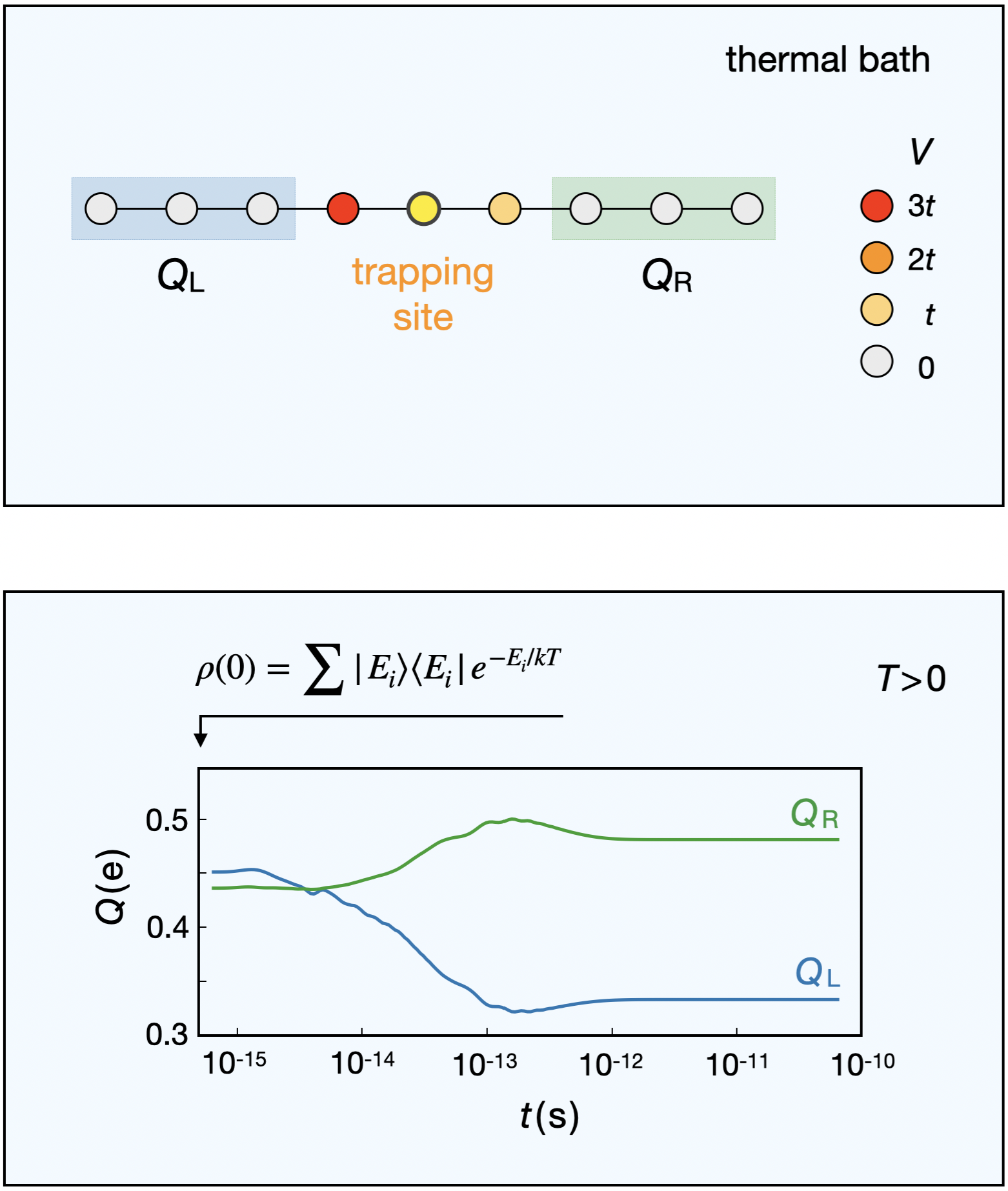}
  \caption{} % leave empty if you want only (a) without text
  \label{fig:Fig_GG-A}
\end{subfigure}\hfill
%\vspace{4em} % adjust vertical spacing between panels
\begin{subfigure}[t]{0.48\textwidth}
  \centering
  \includegraphics[width=\textwidth]{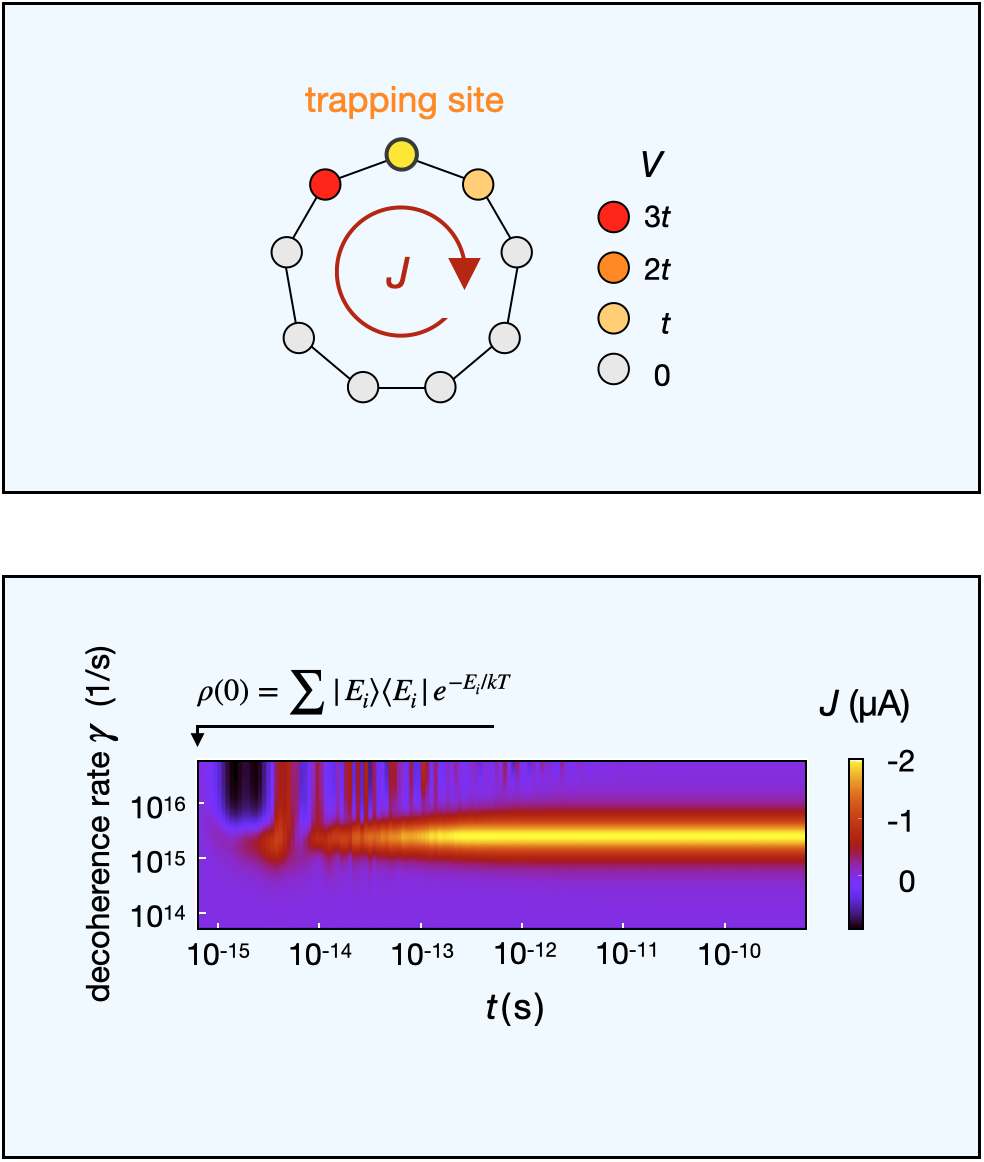}
  \caption{}
  \label{fig:Fig_GG-B}
\end{subfigure}
\vspace{3 mm}
\caption{\textbf{Charge and current dynamics induced by unitary–projective evolution in a minimal tight-binding system.} Figure adapted from \cite{Bredol2021b}.\\
\textbf{(A)} Top: Schematic of the nine-site chain. Bottom: Time evolution of the charges, measured in units of the elementary charge, on the left and right contacts ($Q_L$ and $Q_R$) of the nine-site chain. 
These dynamics were obtained from a Lindblad master-equation calculation within a tight-binding representation of the system \cite{Bredol2021b}. The chain consists of three sites on each side which form the contacts, as well as three central sites with site-dependent electrostatic potentials $V = 3\,t$, $2\,t$, and $t$, from left to right. Here $t$ denotes the hopping amplitude. The central site is coupled to a trap that can absorb an electron and subsequently reemit it as a wave packet. The system is initialized in a thermal density matrix at $k_{\mathrm{B}}T = 1\,\mathrm{eV}$ constructed within the Hamiltonian framework (see Fig.\,\ref{fig:Fig_B}(A)). 
This Hamiltonian thermal ensemble is not a steady state of the unitary–projective dynamics. After a transient period of order $10^{-12}\,\mathrm{s}$, the system relaxes to a new charge-separated steady state. The initial charge imbalance originates from the asymmetric electrostatic potential. The nonreciprocal projection dynamics subsequently reverse it, resulting in unequal electrochemical potentials at the two contacts.\\
\textbf{(B)} Time evolution of the total current in the same system when the contacts are joined to form a ring. The system now relaxes to a non-equilibrium steady state characterized by a circulating current. The bottom panel shows the steady-state current $J$ as a function of time $t$ and wave-packet generation rate $\gamma$. A finite circulating current is obtained for $\gamma \approx 5 \times 10^{14}$–$2 \times 10^{16}\,\mathrm{Hz}$, with $\gamma \approx 2 \times 10^{15}\,\mathrm{Hz}$ corresponding to approximately one inelastic scattering event per electron loop cycle. }

\label{fig:Fig_GG}
\end{figure}

To study the behavior of a system operating in thermal equilibrium, we performed numerical studies based on a Lindblad master-equation approach \cite{Bredol2021b}. To keep the calculations tractable, we employed a minimal model with a small number of tight-binding sites. As shown in the top panel of Fig.\,\ref{fig:Fig_GG}(A), the system consists of a chain of nine sites, with the three central sites subject to on-site potentials of $3\,t$, $2\,t$, and $t$, respectively, where $t$ denotes the nearest-neighbor hopping amplitude. The central site acts as a trap state.

The Lindblad operators were heuristically chosen such that re-emission from the trap populates either left- or right-moving modes. This effectively models a recoil process that stochastically directs the emitted wave packet towards either contact, consistent with the Johnson–Nyquist noise picture. This construction naturally leads to non-Hermitian Lindblad operators. It is not based, however, on a comprehensive microscopic model.

The system is initially prepared in a thermal density matrix corresponding to a single electron at temperature $T$. The trapping probability $p$ is treated as a free parameter. The results, summarized in the bottom panel of Fig.\,\ref{fig:Fig_GG}(A), show the time evolution of the occupations of the left and right contacts. At $t=0$, the Hamiltonian thermal state exhibits slightly different occupations of the two contacts, reflecting the symmetry-breaking potentials of the central sites.

The key observation is that this Hamiltonian thermal state does not remain stationary under the ensuing dynamics. 
Because the asymmetric unitary dynamics causes electrons propagating in opposite directions to encounter the trap with different rates, the otherwise symmetric projection processes act on the transport channels at direction-dependent rates. Consequently, the combined unitary–projective dynamics does not obey detailed balance with respect to the thermal equilibrium distribution defined by the Hamiltonian.

Once the Lindblad evolution is activated, the initial thermal state therefore decays. The contact-occupancy difference first reverses sign and then grows in magnitude, such that the behavior observed over the accessible simulation times strongly suggests convergence towards a non-equilibrium steady state. In this steady state, the system sustains a finite voltage difference between its contacts.

For the sake of completeness, we note that if the system is considered at any intermediate time during its evolution, for example, at $t=2\times10^{-14}\,\mathrm{s}$, its subsequent dynamics will not relax towards thermodynamic equilibrium, as would be expected for a conventional system which obeys detailed balance. Rather, it continues to evolve towards the non-equilibrium steady state, which acts as a dynamical attractor of the reduced dynamics.

From the perspective of statistical physics, a persistent difference in contact occupations corresponds to unequal microstate occupation probabilities.
In thermodynamic equilibrium, such sustained imbalances are forbidden in the absence of thermodynamic forces conjugate to the imbalance. Instead, equilibrium requires a stationary probability distribution that is (i) determined solely by the conserved quantities and (ii) satisfies detailed balance. This precludes the stationary probability currents required to maintain a finite contact-occupation imbalance.

In the present case, however, thermodynamic equilibrium is dynamically inaccessible. Consequently, the system does not relax towards an equilibrium distribution but instead sustains unequal microstate occupation probabilities as an intrinsic feature of its non-equilibrium dynamics. Notably, this  imbalance in occupation arises without any active feedback control involving sensing of particle positions or explicit information processing that conditions the dynamics on measurement outcomes \cite{Lutz2015}. This enables intrinsic material and device properties that originate from persistent imbalances in microstate occupations, or from the dynamics through which the system evolves towards such imbalanced configurations.

Through the charge separation, the electronic subsystem becomes locally more ordered. However, the second law of thermodynamics requires the total entropy of the subsystem plus environment to increase. According to the second law, any local reduction in electronic entropy must be compensated for by greater entropy production in the environment through projection events and dissipative channels.

The voltage generated between the two contacts reflects a difference in the electrochemical potentials or equivalently, a gain of free energy, generated by the u-p dynamics. Therefore, it can, in principle, be used to charge a capacitor, for example. An important question is whether such charging could be performed repeatedly, or even indefinitely, and if not, what the ultimate microscopic physical mechanism is that halts the process. 
The second law is a thermodynamic principle of macroscopic behavior. It does not specify the microscopic forces affecting individual electron trajectories and preventing or halting charge separation. Therefore, establishing whether any such mechanism exists and identifying it, if present, is an important experimental challenge.

When the contacts of such a device are connected to form a closed loop, the generated imbalance drives a circulating current. The corresponding device geometry is shown in the upper panel of Fig.\,\ref{fig:Fig_GG}(B), and the resulting behavior is presented in the lower panel. As the figure illustrates, the system supports a persistent circulating current in its non-equilibrium steady state, even though the current continuously traverses the trap site, where inelastic scattering occurs. This circulating current generates a magnetic moment and an associated magnetic field. 

Because the current is sustained by thermally activated detrapping processes encoded in the Lindblad projection operators, the effect requires finite temperatures. However, these temperatures must be limited, since coherence is lost at sufficiently high temperatures, which suppresses the directional unitary effects, causing the circulating current and its associated magnetic moment to vanish.

The existence and magnitude of the calculated current depend on the specific Lindblad operators employed in the model \cite{Bredol2021b}. Within the present modeling framework, a finite steady-state current only arises when the dissipators have a non-Hermitian structure that incorporates irreversible projection processes into the reduced dynamics. Therefore, a fully microscopic treatment of the underlying trapping and detrapping processes may modify or, in some parameter regimes, even suppress this behavior.

According to the calculations, the resulting currents and magnetic moments are modest but non-negligible. For the considered systems, the maximum currents of order $1 \, \mu \textrm{A}$ correspond to magnetic moments of order $10^{-1}$ Bohr magnetons for a ring diameter of order $1\,\textrm{nm}$. 
Notably, the geometry of the device fixes the sense of the circulating current and, therefore, the polarity of the magnetic moment in a zero applied field. This is in contrast to the spontaneous symmetry breaking that underlies magnetic moments in conventional magnetic materials.

We conclude that materials engaging in the described u-p dynamics can exhibit a novel category of non-equilibrium magnetism. In the considered model, the magnetic moment displays a pronounced maximum as a function of temperature. This magnetism is generated by thermally activated dynamics rather than ground-state mechanisms. This characteristic temperature dependence also does not originate from van Vleck–type mechanisms \cite{VanVleck1932}, but rather from the interplay between the thermal activation of projection processes and thermally induced loss of coherence.

These thermally driven circulating currents may appear counterintuitive at first, but related behavior was identified by Entin-Wohlman et al. \cite{Entin1995, Entin2004}. Using non-equilibrium Green’s-function (Keldysh)–based calculations, they discovered that phonon-assisted processes in magnetic-field-biased mesoscopic rings can generate a thermally induced countercurrent which flows in the opposite direction of the equilibrium persistent current. This countercurrent originates from doubly resonant electron–phonon processes, as first described by \cite{Holstein1961}. Entin-Wohlman et al. emphasized that this constitutes a breakdown of detailed balance in thermal equilibrium \cite{Entin1995, Entin2004}. Using our definition of thermodynamic equilibrium (see the Introduction), the absence of detailed balance indicates that the corresponding stationary state falls outside thermodynamic equilibrium and may be viewed as a non-equilibrium steady state, even though it is maintained by thermal reservoirs.

The persistent current in the present study is analogous to the thermally induced countercurrent discussed above, but differs fundamentally in its origin. It requires no applied magnetic field and instead arises from thermally driven unitary–projective dynamics that stabilizes a non-equilibrium steady state.

\section{Discussion}

In the following, we attempt to develop a unified perspective on unitary–projective materials and devices. We discuss how their behavior emerges from the interplay of coherent Hamiltonian dynamics and irreversible projection processes, how this interplay circumvents the constraints of Hamiltonian transport theory and equilibrium thermodynamics, and how u-p materials relate to concepts such as quantum trajectories, information flow, and non-equilibrium steady states. The present work does not, however, propose a complete thermodynamic theory of unitary–projective systems, but rather identifies robust mechanisms and delineates the regimes in which conventional constraints cease to apply.

\subsection{Context for Unitary–Projective Evolutions}
Unitary–projective materials interface with several broader areas of contemporary quantum science. 
The dynamics of the electronic systems in such materials can be naturally formulated within the framework of open quantum systems, for example using the Gorini–Kossakowski–Sudarshan–Lind\-blad master equation \cite{Gorini1976, Lindblad1976} (hereafter referred to as the Lindblad equation), its stochastic unravelings in terms of quantum trajectories \cite{Plenio1998, Dalibard1992, Molmer1993}, or non-equilibrium Green’s-function techniques \cite{Keldysh1965, Haug2008}.
Parallels further arise in the field of measurement-induced phases, where the interplay of unitary evolution and stochastic projections gives rise to phenomena such as entanglement transitions and non-Hermitian critical behavior \cite{Li2018, Skinner2019, Gullans2020}. 

Between projection events, the deterministic evolution of the system may be captured by effective non-Hermitian Hamiltonians, linking unitary–projective materials to the rapidly developing field of non-Hermitian quantum matter. This framework encompasses phenomena such as nonreciprocal transport and exceptional-point physics \cite{Bergholtz2021}.

While related in spirit to reservoir engineering (see, e.g., \cite{Breuer2002, Poyatos1996}), where tailored dissipation stabilizes desired quantum states, the unitary–projective dynamics considered here differs in a crucial respect: projection events intermittently re-prepare the system, mapping states that retain pathway-dependent correlations onto states in which this dynamical information has been erased. Counter-intuitively, it is precisely this discarding of pathway information—which locally enforces direction-independence—that disrupts the forward–reverse linkage imposed by microreversible unitary evolution and thereby breaks the symmetry between left-to-right and right-to-left transition probabilities in the effective reduced dynamics.

The intrinsic irreversibility and information flow associated with projection events further place u-p materials in the context of quantum thermodynamics and information-powered engines, where measurements act as a resource capable of driving persistent currents or directed responses \cite{Parrondo2015, Sagawa2008, Lutz2015}. A related classical analog is found in ratchet physics, where broken spatial symmetries combined with stochastic driving generate directed motion in the absence of net forces \cite{Reimann2002, Astumian2002}. Unitary–projective materials may be viewed as a specific quantum realization of this idea, in which projection-induced state resets and thermal excitations jointly give rise to non-equilibrium noise modulation acting on quantum states.

The u-p framework also naturally aligns with recent developments in dissipative time crystals \cite{Gong2018}, as projection-stabilized dynamics may provide a general route to dynamical phases beyond equilibrium.

Ideas closely related to u-p systems have been discussed previously in the context of nonreciprocal phase-filtering devices, where direction-dependent phase erasure has been proposed as a mechanism for asymmetric transport and interference phenomena \cite{Mannhart2019b}.
This work examines whether environment-induced information erasure at the system level—occurring without explicit work input for information processing—can enable Maxwell-demon–like behavior, an issue that we revisit explicitly in the Summary.

Collectively, these perspectives define a broad and active theoretical landscape for u-p materials. Next, we examine how their defining properties emerge from the underlying dynamical principles.

\subsection{Irreversible Information Updates in Unitary–Projective Materials in View of Landauer’s Principle}
\label{subsection:info_updates}

A defining characteristic of unitary--projective materials is the role played by information updates in their dynamical evolution. Unlike conventional Hamiltonian quantum materials, whose effective behavior can be understood without explicit reference to information flow, u-p materials operate through projection processes that modify the information content of the electronic subsystem. This modification of information is not a secondary effect, but an essential ingredient that underlies steady states, transport properties, and emergent material responses.

In decoherence-based descriptions of u-p dynamics, projection events irreversibly transfer information from the electronic subsystem to environmental degrees of freedom. This includes, for example, phase relations and information about electron momenta and energies prior to trapping. Conversely, thermally activated release events inject qualitatively different, stochastic information back into the subsystem through sampling of energy, momentum, and phase space. This injected stochastic information redistributes the occupancies among symmetry-broken unitary transport channels.

By contrast, under purely Hamiltonian and microreversible dynamics, including arbitrary unitary backscattering, channel populations evolve coherently and reversibly. Such dynamics does not generate irreversible, internally driven population currents between symmetry-related channels.

The information injected from the environment originates stochastically from state reduction processes. It carries no information from the system’s past and is generated without any explicit work input for information acquisition or processing. Nevertheless, it is neither irrelevant nor meaningless. Once encoded in the system, the information acquires thermodynamic relevance and actively shapes the subsequent state evolution. 
In this way, natural stochastic processes continuously generate a non-exhaustible thermodynamic resource that the system uses to direct its dynamics. This resource becomes externally accessible through the resulting device functionality, without requiring explicit work input associated with controlled information processing.

From the perspective of non-equilibrium thermodynamics, the sustained unitary–projective bias can be viewed as a generalized affinity: a parameter of the reduced dynamics that continuously breaks detailed balance and sustains steady currents, without being consumed by operation. In this sense, u-p systems are driven by dynamical affinities rather than by finite, exhaustible thermodynamic resources.

For clarity, we here distinguish between (i) exhaustible entropic resources stored in non-equilibrium states (e.g., correlations) and (ii) non-exhaustible dynamical resources encoded in the reduced evolution law via sustained projection processes.

At first glance, this behavior appears to contradict Landauer’s erasure principle \cite{Landauer1961}. In his seminal work, Rolf Landauer discusses three classes of memory devices relevant to digital computation. He demonstrated that in those memory devices he analyzed, the logical reset of a bit to a predefined state (in later work referred to as 'discarding information' \cite{Landauer1991}) entails a minimum heat dissipation of $k_{\mathrm B}T\ln 2$. 

This principle he identified is often invoked in a far more universal sense, extending beyond the specific classes of classical devices and the reset operation for which it was originally justified \cite{Landauer1961}. It is commonly taken to imply that any logically irreversible operation, such as erasing any kind of memory bit, must be accompanied by at least $k_{\mathrm B}\ln 2$ of entropy production in the environment, implying an unavoidable minimum heat dissipation of $k_{\mathrm B}T\ln 2$ per bit erased or otherwise irreversibly processed. 

At this point, it is important to recall the physical mechanism underlying Landauer’s derivation of his principle. The minimal dissipation he found arises from the implementation of the reset operation, not from logical irreversibility itself. He formulated his argument within the framework of classical statistical mechanics. 
The devices he used to define the bound are members of his class-I devices, which encode information in bistable memory states. These states are typically realized as electrons residing in one of the wells of a double-well potential. 
This occupancy of these states must be protected against thermal fluctuations on the scale of $ k_{\mathrm{B}}T $.
The minimal energetic cost identified by Landauer arises from the work required to enforce the necessarily dissipative reset operation to the prescribed standard state, corresponding to a many-to-one mapping of logical states, on such stabilized memory elements. 

In contrast, in unitary–projective devices, coherently evolving quantum states carry the relevant information during the unitary segments of the trajectories. Their possible environment-induced decoherence or stochastic interruption is an intrinsic component of the u-p mechanism.
Because this information is not stored in thermally protected configurations but rather transiently during coherent evolution, it does not require energetic stabilization against thermal noise at the system level, as is the case for Landauer’s class-I devices. Nor does it require continuous energy dissipation to maintain stored information, as is necessary for Landauer’s class-II or class-III devices. 
 
Importantly, even the macroscopic imbalances generated by u-p dynamics, such as charge or population differences between contacts, do not constitute memory states that need to be stabilized against fluctuations.
These imbalances are continuously exposed to thermal fluctuations, and thermal excitations in the contacts merely feed particles back into the u-p region, where the same sorting dynamics re-establishes the imbalance.
The imbalance is therefore maintained as a non-equilibrium steady-state property of the open dynamics rather than as a stored bit in a double-well memory cell that requires stabilization against
$k_{\mathrm{B}}T$.

Consequently, information replacement by projection in u-p devices is not constrained by Landauer’s erasure principle as formulated in \cite{Landauer1961}. First, u-p devices lie outside the classes of memory devices analyzed in that work. Second, u-p devices do not operate via a logical information reset to a predefined reference state (e.g., a logical 1), which is the process that, according to Landauer, requires damping and the associated dissipation of energy and that has motivated the general notion of "energy loss due to information erasure". 
Instead, projection replaces information about prior trajectories with uncontrolled, stochastic information induced by the environment. Unitary–projective dynamics is continuous and cyclic, yet it does not implement a controlled reset to a fixed reference state.

Accordingly, the projection-induced information updates of the systems considered here lie outside the domain of applicability of Landauer’s bound, rather than violating it. Landauer’s principle does not assert that every input, change, or loss, of information must be accompanied by a heat dissipation of $k_{\mathrm{B}}T\ln 2$. Here, “information loss” refers to loss in the reduced electronic subsystem, i.e., the relevant physical device. For broader discussions of the scope of Landauer's bound, see, e.g., Refs.~\cite{Hemmo2012, Norton2019}. 

Later generalizations of Landauer’s principle (see, e.g., \cite{Maroney2009, Buffoni2024}) extend the bound to further classes of logically irreversible operations. However, they still presuppose  thermodynamically controlled transformations of identifiable information-bearing states. The uncontrolled, environment-induced stochastic replacement of transient trajectory information in unitary–projective dynamics does not satisfy these assumptions.

The generation of random information in the system underlies the nonreciprocal transmission probability achieved by the u-p dynamics. The environment draws random information and uses it to control changes in the occupancies of the transport channels. In doing so, the environment cannot account for, nor compensate for, the original rate asymmetry with which trajectories interacted with it; information about this asymmetry is therefore effectively erased during the projective trapping process.
Consequently, an asymmetric input rate, corresponding to more frequent probing of one propagation direction than the other, is converted into a symmetric output rate, in which the occupancies of both channels increase by the same amount. 
The pairing of direction-dependent depopulation of the transport channels with their direction-agnostic filling yields, in net effect, a directional imbalance in channel occupancies and, consequently, in the left- and right-propagating transmission probabilities. Intriguingly, this mechanism does not require any symmetry breaking in the environmental response itself.

Therefore, in material-relevant, decoherence-based implementations, the environment acts as an information-generating reservoir with internal degrees of freedom that extract and re-inject information into the electronic subsystem, rather than as a passive bath or an external controller. 
A familiar illustration is given by an ohmic resistor coupled to an environment in thermal equilibrium and generating Johnson–Nyquist noise \cite{Johnson1928, Nyquist1928}. The coupling to the environment both dissipates fluctuation energy and gives rise to non-exhausting Johnson–Nyquist noise through stochastic and local equilibrium fluctuations.

In objective collapse models, projection events are not mediated by a conventional environment. Instead, the collapse dynamics removes information associated with coherent superpositions and replaces it with the information corresponding to the realized projection outcome. Although this process does not involve information exchange with a macroscopic bath, it may be understood as transfer to and from additional degrees of freedom, such as the fast variables proposed in emergent-quantum-mechanics frameworks \cite{tHooft2021}. In either case, the electronic subsystem undergoes irreversible information updates, and the resulting reduced dynamics departs fundamentally from purely Hamiltonian evolution.

U-p materials realize a form of information-selective dynamics, in which projection processes—whether environment-mediated or collapse-based—play an active role in shaping electronic behavior. U-p materials can therefore be viewed as physical systems whose functionality emerges from controlled, yet stochastic, information updates, rather than from static band structure alone. 

Now that we have established how projection-induced information updates generate directional transport and non-equilibrium steady states, we will turn to the constraints that govern unitary-projective dynamics. We will also discuss the consequences of these dynamics for charge, current, and magnetic responses.

\subsection{Constraints and Consequences of Unitary–Projective Dynamics}

Compared to conventional Hamiltonian systems, which are constrained by microscopic reversibility and the associated reciprocity and linear-response relations (see Sect.~\ref{sec:Evolution_of_States}), u-p materials are subject to fewer restrictions. However, for the electronic subsystem the dynamics must still respect global conservation laws and symmetries. These requirements limit the amount of transport directionality and free energy that projective dynamics can generate, even though reciprocity and linear-response relations need not apply. An analog of the Onsager–Kubo–Landauer framework for unitary–projective systems that specifies the corresponding constraints has yet to be developed.

A further fundamental point is that projection-induced state updates directly modify the spatial distribution of charge and the current flow. Charge density governs virtually all electronic properties of a material, while circulating currents determine its magnetic response. As the two examples above demonstrate, u-p dynamics can reshape both. Therefore, u-p evolution directly tailors a material's most basic electronic and magnetic characteristics.

\subsection{Beyond Effective Hamiltonian Descriptions}

The functionalities enabled by u-p dynamics cannot be reproduced by any effective Hermitian or non-Hermitian Hamiltonian description in the absence of measurement backaction. Even when augmented by disorder, interactions, or non-Hermitian terms, Hamiltonian evolution remains linear and microreversible in terms of probability flow unless explicitly biased or driven. This remains true even for quantum materials that exhibit nonreciprocal responses within a Hamiltonian framework, such as non-centrosymmetric systems with broken inversion or time-reversal symmetry \cite{Tokura2018, Nagaosa2024}.

\subsection{Characteristic Temperature Dependence}
\label{subsec:T_peak}

The requirement for an electron to undergo, on average, say, between $10^{-2}$ and $10^{2}$ inelastic scattering events while otherwise proceeding ballistically during its passage through a microscopic unit leads to a central characteristic of u-p materials: a distinctive temperature dependence of their macroscopic properties. These typically exhibit a pronounced maximum, arising from the competition between thermally activated projection processes required for u-p operation and the thermally induced loss of coherence at higher temperatures.

One particularly direct way to probe such u-p electronic dynamics experimentally is to measure this temperature dependence in ordered assemblies of asymmetric cyclic organic molecules. In these systems, the magnetic moment generated by u-p dynamics is expected to exhibit a pronounced maximum at an intermediate temperature, accompanied by a corresponding peak in magnetic susceptibility.

\subsection{Scales}

Length and energy scales are key parameters that govern quantum effects, often confining practical applications to small devices operated in shielded environments at low temperatures in order to preserve coherence. In contrast, u-p effects require operation in a regime of neither fully coherent nor fully diffusive transport, characterized by a small but finite number of inelastic scattering events per electron passage. This condition can be satisfied at room temperature in small organic molecules. In such systems, the characteristic energy scales associated with microscopic transport are set by energies of the order of $k_{\mathrm{B}}T$ at $300\,\mathrm{K}$.

Macroscopic architectures can, in principle, be realized by incoherently coupling large numbers of molecular units into extended networks  (Fig.\,\ref{fig:Fig_Mac}). Their resulting response is expected to scale extensively, potentially enabling macroscopic systems to operate at room temperature or even above with substantial practical value.

An important open question concerns the timescale over which the functionality of a unitary–projective system, for instance, capacitor charging, might decay due to operation-induced environmental backaction that modifies the effective projection processes.  

As will be discussed in Sect.~\ref{subsec:enhanced_space}, u-p dynamics is not generally formulated in terms of consuming a finite, exhaustible environmental resource. 
In many relevant classes of u-p implementations, the net energy, momentum, or phase transferred to the environment has a mean value of zero, so that the environment experiences no systematic drift and remains effectively unchanged at the coarse-grained level. In such cases, device performance is not expected to degrade through cumulative bath depletion.

If a finite exhaustion nevertheless arises, for example due to small systematic imbalances, slow environmental modes, or imperfect cancellation, standard scaling considerations show that coupling the system to an environment with sufficiently large effective capacity can make the associated degradation timescale arbitrarily long, ensuring that operation-induced changes remain negligible over the timescale of interest.

\subsection{Electron Wave Packets In and Near Thermodynamic Equilibrium}

One might argue that an electronic system exhibiting well-defined, spatially confined wave packets is far from thermal equilibrium. However, this argument is only valid if thermodynamic equilibrium is identified exclusively with a static, spatially homogeneous thermal density matrix, without taking into account localized degrees of freedom or microscopic emission processes. This picture must be modified when trap-like defects are present. 

In thermal equilibrium, a trap state is occupied only intermittently, and whenever an electron leaves the trap, it is emitted as a spatially confined wave packet. 
This process can be compared to the Wigner–Weisskopf decay of a discrete, excited atomic or nuclear state coupled to a continuum \cite{Weisskopf1930}.
Over the subsequent coherence time and coherence length of the system, the wave packet propagates away from the defect, scatters, disperses, and eventually loses its spatial confinement.
Therefore, even in or near thermal equilibrium, defects continually generate electronic wave packets in their vicinity within a spatial region set by the coherence length.  

In the solid-state setting considered here, detrapping occurs into an excited state and is typically induced by phonon absorption. Phonon-induced detrapping constitutes a stochastic state-preparation process because the absorbed phonon provides both energy and momentum as drawn from an incoherent thermal environment.

As a result, electronic states with potentially complex and stochastic spatial structures that impinge on a trap are stochastically replaced by outgoing wave packets with different structures. The properties of these emitted wave packets, which remain coherent over the coherence length, are governed by the trap’s thermal emission process and the absorbed phonon. The molecules, unit cells, and devices discussed in this manuscript operate on precisely this mesoscopic scale, where the emitted wave packets retain quantum coherence.

While wave-packet emission events are fully permitted in thermal equilibrium, the ensuing unitary–projective dynamics can, in the presence of asymmetric unitary propagation, drive the electronic subsystem away from equilibrium and towards a non-equilibrium steady state.  In this state, wave packets remain the natural carriers of the dynamics.

\subsection{Why Global Unitarity Does Not Imply Thermodynamic Equilibrium}

It is also worth clarifying that the mixed unitary–projective evolution of an electronic subsystem coupled to an environment can, in principle, be embedded within the evolution of a larger system that is fully unitary. However, this does not imply that the environment acts as a passive equilibrium reservoir or that the electronic subsystem must operate in thermodynamic equilibrium.

Instead, the projection processes considered here endow the environment with an effectively active role that goes beyond that of a passive thermal reservoir: it extracts information from the electronic subsystem and induces irreversible state updates in its reduced dynamics.
In such a setting, the usual expectation that the combined system–environment relaxes towards a global thermodynamic equilibrium no longer applies. For u-p systems, even if the total system evolves unitarily, tracing out the environment yields reduced dynamics that are generically stochastic and nonunitary, violating detailed balance unless the environment acts as a passive thermal reservoir. Consequently, the electronic subsystem does not relax to a thermal state but instead approaches a non-equilibrium steady state.

Moreover, when the subsystem is externally prepared in a configuration corresponding to standard thermodynamic equilibrium, this equilibrium state is dynamically unstable: the ensuing unitary–projective dynamics drives the system away from equilibrium towards a non-equilibrium steady state that permits and sustains unequal occupation probabilities of microstates (see Sec.\,\ref{subsection:novel_magnetism}). As a result, currents, voltages, or population imbalances can, in principle, persist indefinitely.

\subsection{Tensions Between the Born Rule and the Second Law of Thermodynamics}
\label{subsec:second_law}

The dynamics arising from the combined unitary and projective evolution delineated above reveals fundamental tensions within standard theoretical frameworks and challenges the second law of thermodynamics (see, e.g., \cite{Mannhart2018, Bredol2021b, Braak2020}). 

It is important to distinguish this type of tension from other lines of work in quantum thermodynamics that extend the standard formulation of the second law by incorporating additional physical ingredients. 
In particular, generalized frameworks that account for correlations, non-equilibrium reservoir states, or strong system–environment coupling modify the conventional entropy balance by incorporating additional terms, while preserving the validity of the second law within an enlarged description (see, e.g., \cite{Parrondo2015}). 
In such approaches, apparent violations of textbook bounds arise because the standard formulation neglects relevant degrees of freedom, not because the second law itself breaks down.

By contrast, the tensions discussed here arise already at the level of the reduced dynamics constructed from unitary Hamiltonian evolution supplemented by Born-rule transition probabilities. 
They therefore point to a more fundamental incompatibility between the intrinsically asymmetric projection processes and the conventional assumptions underlying the second law, rather than to a mere incompleteness of thermodynamic bookkeeping. 
In this sense, the examples discussed below do not simply motivate generalized entropy balances, but instead raise the question of whether the compatibility between the Born rule and the second law can be taken for granted in the presence of asymmetric projection dynamics.

A striking example is provided by the thought experiment presented in \cite{Braak2020}, which considers ensembles of two-level systems coupled nonreciprocally to chiral photonic waveguides and to black-body reservoirs. The transition processes of the two-level systems are treated in the standard way using Fermi’s golden rule, which is derived from the Born rule in the weak-coupling limit for transitions into a continuum of final states (see Sect.~\ref{subsec:Projection_dynamics}). 
Significantly, when the two-level system's chirality-dependent couplings to the waveguides are combined with such probabilistic radiation processes, the resulting rate equations violate detailed balance. This leads to negative entropy production over finite time intervals, in apparent contradiction with Clausius’ formulation of the second law of thermodynamics.

This behavior originates from the interplay between nonreciprocal unitary dynamics and stochastic projection processes in the reduced electronic system.  As in the other examples discussed above, unitary states related by a symmetry, in this case, the chiral propagation modes of the waveguides, are stochastically intermixed by the environment (here, black-body reservoirs), with mixing rates that depend on the unitary states themselves via state-dependent coupling constants. The analysis therefore reveals that the commonly unquestioned compatibility between Born-rule transition probabilities, a cornerstone of quantum mechanics, and the second law of thermodynamics need not hold when projection events mix the occupations of symmetry-broken transport channels.

\section{Outlook}

\subsection{Enhanced Functional Space}
\label{subsec:enhanced_space}

Together, unitary evolution and projection processes can realize a broad class of quantum state transformations that extend far beyond those accessible under closed, purely unitary dynamics. The design freedom on the unitary side spans the full range of complexities familiar from electronic-structure theory, including strong electronic correlations, topological effects, and interface and heterostructure engineering. The projective component, by contrast, introduces an orthogonal design axis that remains largely unexplored.

It is therefore plausible that, in non-equilibrium settings where microreversibility and detailed balance are broken, unitary–projective materials may realize functionalities or phases of matter beyond the single-particle valves, persistent currents, voltages, and non-equilibrium magnetism demonstrated here. These include thermoelectric responses outside the scope of Onsager theory, noise-driven rectification effects, lossless charge transport \cite{Mannhart2018, Mannhart2019a}, and measurement-stabilized non-equilibrium phases.

Moreover, unitary–projective dynamics introduces qualitatively new modes of materials control. 
Fundamental material properties such as electric polarization or magnetic moments can be tuned not only through conventional Hamiltonian parameters, but also through the controlled manipulation of projection processes, for instance by external electric or magnetic fields. This dual control mechanism has no analog in materials designed using conventional electronic-structure theory.

An important question regarding this enhanced functional space concerns the thermoelectric performance and the thermodynamic classification of unitary–projective materials and devices. In particular, their operation challenges the applicability of standard heat-engine benchmarks and raises questions about how performance limits should be formulated for this broader class of systems.

If unitary–projective systems are operated as thermal machines, they can generate electrochemical potential differences even under isothermal conditions \cite{Bredol2021b}. When such systems are naively treated as standard thermal engines, their apparent efficiency therefore exceeds the conventional Carnot limit. In fact, it may become formally unbounded when temperature differences vanish. This reflects a misapplication of the traditional Carnot bound rather than a genuine thermodynamic violation.
The Carnot efficiency only applies to heat engines that operate reversibly between thermal reservoirs in equilibrium and that are driven exclusively by temperature differences as the thermodynamic resource. Unitary–projective systems fall outside this class. Consequently, assessing their performance in thermodynamic terms requires explicitly accounting for the energetic and entropic costs of projection-induced state updates.

One possible approach is to retain a Carnot-type interpretation by considering the sustained non-equilibrium bias generated by projection processes as an additional thermodynamic resource that can supplement or replace a temperature difference. 
From this perspective, the performance of u-p devices can be described using generalized heat-engine frameworks that incorporate as resources entropic contributions beyond heat. A recent, analogous formulation of a corresponding generalized Carnot framework, applicable to systems that exploit system–environment correlations as a resource, is presented in \cite{Aguilar2025}.

A second approach is to view u-p systems as not being Carnot-type engines at all,  but rather as a distinct class of non-thermal machines operating outside the heat-engine paradigm. In unitary–projective systems, the non-equilibrium bias induced by projection processes enters directly into the dynamical laws governing the reduced system, rather than being encoded as a finite, exhaustible resource stored in the state. Therefore, their steady states need not correspond to thermodynamic equilibrium. Detailed balance is generically broken, and transport is driven by non-thermal dynamical biases, instead of, or in addition to, temperature differences \cite{Mannhart2018, Bredol2021b, Mannhart2021}. From this perspective, efficiencies defined with respect to heat absorbed from thermal reservoirs can exceed the Carnot limit. This is not because additional entropic fuel is consumed by a Carnot-type engine, but because the assumptions underlying Carnot’s bound are simply inapplicable.

Regardless of the chosen approach, no general upper efficiency bounds are currently known for materials and devices governed by unitary–projective dynamics. Clarifying whether fundamental limits exist in this broader class of non-thermal machines, and determining how they are set by dynamical, informational, or stability constraints, remains an open and promising avenue for future research.

The dynamical inaccessibility of thermodynamic equilibrium substantially expands the space of accessible material functionalities.
These functionalities can be realized in individual devices, engineered metamaterials, or even bulk materials. Potential applications range from high-sensitivity sensing to quantum information processing and, importantly, energy harvesting and conversion.

\subsection{Theoretical Directions}

Generalized formulations of the second law that explicitly account for system–environment correlations and information flow have recently been advanced (see, e.g., \cite{Parrondo2015}). However, a precise and quantitatively predictive theory of the open-system dynamics underlying u-p materials is still lacking. At present, no comprehensive framework exists that can treat such systems in a fully microscopic and predictive manner. Developing such a theory, which must also incorporate many-body effects, is essential both for rational materials design and for achieving quantitative agreement between theoretical predictions and experimental observations.

In particular, a key open challenge is formulating theoretical models that describe projection events in real materials at a microscopic level. It would be highly desirable, for example, to treat the trapping and detrapping processes discussed in this work with microscopic accuracy within an explicit open-environment setting, rather than relying on phenomenological projection operators or effective rates. 

Because u-p functionality emerges from the interplay of coherent propagation, symmetry breaking, and projection-induced state updates, materials design in this context amounts to identifying the appropriate Hamiltonians, geometries, and projection mechanisms that, when combined, produce the desired dynamical response. Another task is understanding how information flow, entropy production, and resource constraints set fundamental performance bounds on u-p functionalities.

Owing to the high dimensionality and intrinsic nonlinearity of this design space, data-driven and AI-based approaches are expected to be valuable tools for this endeavor. 
Methods that autonomously explore parameter spaces by repeatedly solving quantum-trajectory equations, Lindblad-type master equations, or related open-system models can reveal non-intuitive combinations of unitary pathways and projective channels that lead to robust u-p behavior. Developing such inverse-design strategies, grounded in physically transparent dynamical models, represents a promising direction for future condensed-matter theory.

To enable rational design and physical intuition for measurement-driven devices, a new conceptual tool analogous to band diagrams is furthermore needed. This tool should provide visually intuitive and quantitative access to the underlying measurement-driven, asymmetry-based processes, rather than to static energy levels alone.

Beyond their practical relevance, these efforts also involve questions of fundamental importance. The behavior of u-p systems may challenge aspects of established thermodynamic principles: to date, no explicit microscopic mechanism has been identified that would guarantee positive entropy production or preclude continuous work extraction in such dynamics. Indefinite charging cycles enabled by u-p valves operating in their non-equilibrium steady states (see Sec.\,\ref{subsection:novel_magnetism}) would, if permitted by nature, stand in contradiction to the second law of thermodynamics in the sense of Kelvin’s formulation. Crucially, as described in Sec.,\ref{subsec:second_law}, theoretical studies have revealed a significant discrepancy between the predictions of the second law of thermodynamics and those obtained from standard golden-rule modeling of specific chiral, nonreciprocal systems \cite{Braak2020}.

In this context, exploring the behavior of deliberately inhomogeneous devices and materials will also be useful. For instance, the countercurrent scenario of Refs.\,\cite{Entin1995, Entin2004}  can be applied to rings where a small sector experiences additional weak conventional inelastic electron-phonon scattering in the steady state. Similarly, a sector with enhanced conventional inelastic scattering can be introduced into rings with persistent currents in non-equilibrium steady states, or into regions supporting persistent edge currents in quantum Hall systems or related topological structures. A microscopic analysis of the resulting spatial distribution of charge, dissipation, and temperature is important and worthwhile.

This tension underscores the necessity of carefully designed experiments to determine whether analogous inconsistencies emerge, or are resolved, in physical implementations of unitary–projective materials and devices operating beyond the constraints of the Onsager–Kubo–Landauer-\-B\"uttiker regime. Such experiments may even offer new insights into the nature of the quantum-mechanical measurement process itself and into the dynamics underlying projection events.

\subsection{Materials Directions}

Materials whose electronic behavior is governed by u-p evolutions must contain unit cells or mole\-cules that support sufficiently long scattering lengths to allow predominantly unitary transport, interrupted occasionally by projection events. These building blocks can be assembled into macroscopic structures by connecting the molecular or crystalline units via incoherent contacts. The overall dimensions of these structures may greatly exceed the scattering length.

Organic molecules are particularly versatile building blocks, offering a favorable combination of coherence length, structural asymmetry, and tunable dissipation pathways. Prominent examples with appropriately broken symmetries include asymmetric aromatic rings coupled to contacts, conjugated organic molecules with asymmetric backbones, and molecules bearing side groups arranged in asymmetric substitution patterns (see Fig.\,\ref{fig:Fig_M} for examples). Such molecules can be interconnected into two- or three-dimensional ensembles to create macroscopic systems, as exemplarily illustrated in Fig.\,\ref{fig:Fig_Mac}(A). Notably, there are no fundamental obstacles to synthesizing suitable molecules or forming self-assembled monolayers from them. Ensembles featuring molecular-level interconnections, however, require further research and development.

These design principles can also be implemented in inorganic crystals, for example through tailored layering sequences in heterostructures (Fig.\,\ref{fig:Fig_Mac}(B)), asymmetric unit cells in three-dimensional crystalline lattices, metamaterial architectures, and patterned two-dimensional materials. In the latter case, even macroscopic quantum systems, such as quantum Hall edge channels (see \cite{Weis2024} for an overview), may be incorporated as functional elements.

Projection centers in these systems need not be exotic. They can be realized across all these material classes through specific molecular groups, ions, or defects coupled to a continuum of states. Beyond these engineered elements, ubiquitous processes, such as inelastic electron–phonon scattering and strain- or stress-mediated coupling of ionic degrees of freedom to the bulk, naturally produce effective projection dynamics. Moreover, these material platforms can be combined, for example in phase mixtures or heterostructures, enabling additional functionalities that emerge from interfacial coupling and multilayer architectures.

The presence of characteristic temperature-dependent responses in u-p dynamics, such as those discussed in Sect.~\ref{subsec:T_peak}, provides a concrete experimental pathway to identifying and developing suitable material platforms. Beyond engineered systems, the possibility that nature itself may already exploit u-p mechanisms in biochemical settings is worth exploring. The enormous diversity of functional biomolecules may well include a subset for which the room-temperature inelastic mean free path is comparable to molecular dimensions—an essential condition for realizing combined unitary–projective state evolution.

\subsection{Generality of Unitary–Projective Dynamics}

Viewed abstractly, unitary–projective dynamics can be understood as a mechanism by which stoch\-astic processes, irreversible information updates, and symmetry breaking alter the outcomes of otherwise Hamiltonian-governed dynamics of the system. In this way, information about an original symmetry or asymmetry arising from an effectively Hamiltonian evolution is erased, while its dynamical consequences persist. 

For example, an ensemble of skewed, symmetry-broken values of some property, such as impact rates depending on the global direction of trajectories, is passed to a stochastic process that replaces these values by random, unskewed ones. In this example, the impact rates are converted into output rates that are independent of the trajectories' directions. These symmetrized rates then control the system's subsequent evolution. Because this update does not preserve the directional information necessary for a purely Hamiltonian evolution to remain microreversible, the resulting effective reduced dynamics can exhibit different symmetry properties than those implied by the underlying Hamiltonian alone. In particular, it can produce stationary reduced states with symmetry-broken microstate populations.

This symmetry breaking does not arise from a sophisticated—let alone intelligent—control of individual trajectories. Rather, the updating mechanism is trivial, purely stochastic, and unbiased. The asymmetry originates in the prior Hamiltonian dynamics and does not rely on active feedback or intelligent intervention.

Against this backdrop, the device principles discussed in this paper consistently reveal an inherent tension between microscopic dynamics and the second law when particle populations, which evolve under symmetry-broken Hamiltonian dynamics, are reshuffled by intermittent stochastic processes.

More broadly, the principles underlying unitary–projective (u–p) dynamics suggest that their operation is not confined to electronic states in mesoscopic devices. 
One may therefore ask whether analogous mechanisms operate in other areas of science, in which Hamiltonian-type propagation is intermittently interrupted by stochastic or dissipative processes.
Examples include chemical reactions involving interference between competing pathways \cite{Dai2003, Aoiz2018}, aspects of catalytic reaction dynamics, certain processes in particle physics, and, importantly, classes of non-quantum systems.

In non-quantum settings, stochastic behavior could arise, for example, from effective coupling to a thermal bath, to a deterministic but chaotic system, or to a driven non-equilibrium reservoir.
Under such conditions, u--p--type dynamics could generically bias the relative weights of competing decay, transport, or reaction pathways, leading to asymmetric outcomes or shifted steady-state balances.
Identifying and characterizing such systems beyond mesoscopic electronics remains an open challenge.

It is of particular interest to ask whether some aspects of u–p–type behavior may already arise in a broader class of systems that do not even require projection processes. Systems, that do not relax to thermodynamic equilibrium but instead settle into non-equilibrium steady states, break detailed balance, and are structured to harvest the resulting imbalances through coupling to external degrees of freedom.

Importantly, within their stated assumptions, the results presented here point to Maxwell-demon–like behavior \cite{Maxwell1867, Leff2014} in such systems, in the sense that functional asymmetries arise without explicit measurement-based feedback or external control. 
A promising direction for future research, with potentially significant technological implications, is therefore to investigate whether such devices can be physically realized and whether the predicted effects can be confirmed experimentally.

\section{Summary}
\label{sec:summary}

Designing robust and functional unitary–projective dynamics provides a route to materials and devices whose behavior goes beyond the functional range allowed by conventional Hamiltonian-only electronic-structure models. In systems with unitary–projective dynamics, functionality emerges from the interplay of coherent unitary evolution and stochastic projection processes that induce irreversible information updates in the electronic subsystem. 
In this setting, electron dynamics is trajectory-based rather than governed by energy eigenstates, and thermodynamic equilibrium need not be reachable or stable at the level of the electronic subsystem. For the systems considered here, this results in relaxation towards non-equilibrium steady states whenever equilibrium is dynamically inaccessible.

More fundamentally, unitary–projective dynamics enables qualitatively new physical behavior by driving stochastic population transfer between symmetry-defined dynamical sectors of the reduced electronic system (e.g., left- and right-moving transport channels), with symmetry-broken transfer rates. This stochastic coupling is essential for the emergence of new dynamics, steady states, and material properties that cannot be attained within purely unitary evolution of the reduced electronic subsystem.

In particular, electronic dynamics in unitary–projective materials do not need to obey the assumptions underlying equilibrium or near-equilibrium Hamiltonian systems, such as detailed balance, linear-response relations (e.g., Onsager reciprocity and fluctuation–dissipation), or the existence of local thermal equilibrium. They also do not need to obey the thermodynamic bounds derived from these assumptions. In contrast,
in unitary–projective materials the intrinsic information transfer between system and environment constitutes an additional non-equilibrium, non-exhaustible resource.

As a consequence, efficiency bounds, such as the Carnot limit and its generalizations, which are defined for thermal engines operating between equilibrium reservoirs and driven by temperature differences, possibly supplemented by athermal, information-thermodynamic resources, are inapplicable to non-thermal transport machines based on unitary–projective dynamics.

Super-Carnot efficiencies observed in u-p systems reflect operation in a regime where the heat-engine paradigm is not an appropriate conceptual framework. Performance is instead governed by non-equilibrium dynamical biases and sustained probability currents, rather than temperature differences and entropic resources.

For systems governed by unitary–projective dynamics, decoherence and measurement backaction therefore cease to be merely detrimental and instead become functional design elements. 
By converting direction-dependent dwell times into persistent imbalances in microstate occupation, they  alter the system’s steady-state behavior fundamentally. 
Such microstate occupation imbalances are particularly significant because, in thermodynamic equilibrium, microstates consistent with the conserved quantities must be populated according to a fixed stationary distribution, equal in the microcanonical ensemble or Boltzmann-weighted in the canonical ensemble.
This requirement is the fundamental postulate of statistical physics that underlies the exclusion of Maxwell-demon–type behavior \cite{Lutz2015, Leff2014}. 

Sustaining  non-equilibrium microstate occupation imbalances through unitary–projective dynamics in tailored systems provides a direct microscopic cause for persistent currents, voltages, and magnetization. These phenomena are excluded within equilibrium statistical mechanics and under purely unitary dynamics. Importantly, they remain compatible with a description of the full system governed by a Hamiltonian: sustained nonreciprocal responses arise only when the environment participates as a non-equilibrium information reservoir. 

While the phenomenological and model-based approaches employed here reveal consistent and physically motivated behavior, a fully microscopic, first-principles theory of electron dynamics in open quantum systems is still lacking in general. 
In particular, a microscopic treatment of trapping and detrapping processes may, for certain geometries or parameter regimes, restore detailed balance and stabilize thermodynamic equilibrium, thereby suppressing charge separation or persistent macroscopic currents.

Until such a framework is developed or stringent experimental validation becomes available, the present results should be understood as preliminary, mechanism-focused predictions that identify robust physical principles and delineate the conditions under which unitary–projective dynamics can give rise to novel functionalities, rather than as universally applicable, quantitatively definitive, material-specific calculations. 
This paper aims to motivate and guide future theoretical and experimental investigations in this broad and emerging field, where substantial unexplored territory offers many opportunities to discover new physical phenomena.

Looking ahead, unitary–projective materials hold promise both for enabling novel functionalities and for probing basic questions. 
On the materials and device side, by outmaneuvering (i) the constraints associated with closed, near-equilibrium Hamiltonian dynamics and (ii) those of equilibrium thermodynamics, unitary–projective dynamics opens a vast design space for materials with unprecedented functionalities, including platforms for sensing, energy harvesting, and conversion. These platforms could have significant practical impact.

On the theoretical side, u‑p systems provide a testbed for exploring questions concerning the limits of non‑reciprocal behavior—questions that arise at the intersection of quantum measurement, information theory, and thermodynamics. 

Realizing these opportunities will require careful materials engineering and synthesis, thermodynamically consistent modeling of microscopic quantum dynamics, and creative experimental design. The potential payoff, however, comprises new classes of quantum matter and quantum devices that utilize quantum projections as a core mechanism and that may enable us to address problems long considered unsolvable for fundamental reasons.

\vspace{2.5em}

\section*{Materials and Methods}
A large language model (ChatGPT\,5.2) was used for language editing, stylistic refinement, and consistency checks of the manuscript.

\vspace{2.5em}

\section*{Acknowledgments}

The author acknowledges very helpful discussions with A.~Alavi, D.~Braak, T.~Harada, V.~Harbola, S.~Jung, T.~Kopp, R.~Kremer, D.~Manske as well as with many other colleagues. 

\end{document}